\documentclass[letterpaper, onecolumn, unpublished, allowtoday]{quantumarticle}
\pdfoutput=1 

\usepackage[utf8]{inputenc}
\usepackage[english]{babel}
\usepackage[OT2,T1]{fontenc}        
\usepackage{dsfont}                 
\usepackage{bbold}                  
\usepackage[RGB,dvipsnames]{xcolor} 

\definecolor{mygreen}{RGB}{34, 136, 51}
\definecolor{myblue}{RGB}{68, 119, 170}
\definecolor{myred}{RGB}{238, 102, 119}
\definecolor{mypurple}{RGB}{170, 51, 119}
\definecolor{mycyan}{RGB}{102, 204, 238}

\usepackage{comment}                
\usepackage{todonotes}              
\setuptodonotes{inline,
                backgroundcolor=yellow!30,
                prepend,
                caption={\tt}}
\usepackage{lipsum}     
\usepackage[en-US]{datetime2}
\usepackage{setspace}                         
\usepackage{graphicx}                
\usepackage{subcaption} 
\usepackage{float}                   
\usepackage[export]{adjustbox}
\usepackage{array}
\usepackage{multirow}

\usepackage{amsmath}           
\usepackage{amsfonts}          
\usepackage{amssymb}           
\usepackage{algorithm}        
\usepackage{algorithmicx}      
\usepackage[noend]{algpseudocode}
\usepackage[colorlinks=true,
            urlcolor=blue,
            citecolor=blue,
            linkcolor=blue]{hyperref}
\usepackage[square,numbers,comma,sort&compress]{natbib} 
\usepackage{qcircuit}
\usepackage{braket}

\usepackage{silence}
\usepackage[dvipsnames]{xcolor}

\graphicspath{{figures/}}

\usepackage[normalem]{ulem}

\newcommand{\pluseq}{\mathrel{+}=}

\newcommand{\CC}{{\mathds{C}}}

\newcommand{\<}{\langle}
\renewcommand{\>}{\rangle}

\newcommand{\lsim}{\mathrel{\raise.3ex\hbox{$<$\kern-.75em\lower1ex\hbox{$\sim$}}}}
\newcommand{\gsim}{\mathrel{\raise.3ex\hbox{$>$\kern-.75em\lower1ex\hbox{$\sim$}}}}

\def\QECCnk[[#1,#2]]{[\![#1, #2]\!]}

\def\QECCnkq[[#1,#2,#3]]{[\![#1, #2]\!]_{#3}^{\vphantom{T}}}

\def\QECCnkd[[#1,#2,#3]]{[\![#1, #2, #3]\!]}

\def\QECCnkdq[[#1,#2,#3,#4]]{[\![#1, #2, #3]\!]_{#4}^{\vphantom{T}}}

\def\QECCnkgd[[#1,#2,#3,#4]]{[\![#1, #2, #3, #4]\!]}

\def\QECCnkgdq[[#1,#2,#3,#4,#5]]{[\![#1, #2, #3, #4]\!]_{#5}^{\vphantom{T}}}

\def\QECCnkdc[[#1,#2,#3,#4]]{[\![#1, #2, #3; #4]\!]}

\def\QECCnkdcq[[#1,#2,#3,#4,#5]]{[\![#1, #2, #3; #4]\!]_{#5}^{\vphantom{T}}}

\def\QECCnkgdcq[[#1,#2,#3,#4,#5,#6]]{%
  [\![#1, #2, #3, #4; #5]\!]_{#6}^{\vphantom{T}}}

\newcommand{\bigO}{{\cal O}}

\newcommand\CNOT{\ensuremath{\textit{CNOT\/}}}

\newcommand\TOF{\ensuremath{\textit{TOF\/}}}
\newcommand\SWAP{\ensuremath{\textit{SWAP\/}}}

\def\openone{\leavevmode\hbox{\small1\normalsize\kern-.33em1}}

\newcommand{\calB}{{\mathcal{B}}}   
\newcommand{\calC}{{\mathcal{C}}}   
\newcommand{\calP}{{\mathcal{P}}}   

\newcommand{\viz}{\textit{viz.}}
\newcommand{\ie}{\textit{i.e.}}

\newcommand{\eg}{\textit{e.g.}}

\newcommand{\Schroedinger}{{Schr\"{o}dinger}}

\long\def\symbolfootnote[#1]#2{\begingroup%
\def\thefootnote{\fnsymbol{footnote}}\footnote[#1]{#2}\endgroup}

\newlength{\blob}
\newcommand{\sx}[1]{\makebox[\blob][c]{$#1$}}

\newcommand{\cubecode}{\ensuremath{[\![8, 3, 2]\!]}}
\newcommand{\squarecode}{\ensuremath{[\![4, 2, 2]\!]}}

\newcommand\CCZ{\ensuremath{\textit{CCZ\/}}}

\begin{document}

\title{A small and interesting architecture\protect\newline for early fault-tolerant quantum computers}

\author{Jacob S. Nelson}
\affiliation{Quantum New Mexico Institute (QNM-I),
             Center for Quantum Information and Control (CQuIC),\\
             Department of Physics and Astronomy,
             University of New Mexico,
             Albuquerque, NM 87131, USA}
\affiliation{Center for Computing Research (CCR),
             Quantum Algorithms and Applications Collaboratory (QuAAC),\\
             Department of Quantum Computer Science,
             Sandia National Laboratories,
             Albuquerque, NM 87185, USA}
\email{jnelso2@sandia.gov}

\author{Andrew J. Landahl}
\affiliation{Quantum New Mexico Institute (QNM-I),
             Center for Quantum Information and Control (CQuIC),\\
             Department of Physics and Astronomy,
             University of New Mexico,
             Albuquerque, NM 87131, USA}
\affiliation{Center for Computing Research (CCR),
             Quantum Algorithms and Applications Collaboratory (QuAAC),\\
             Department of Quantum Computer Science,
             Sandia National Laboratories,
             Albuquerque, NM 87185, USA}
\email{alandahl@sandia.gov}

\author{Andrew D. Baczewski}
\affiliation{Quantum New Mexico Institute (QNM-I),
             Center for Quantum Information and Control (CQuIC),\\
             Department of Physics and Astronomy,
             University of New Mexico,
             Albuquerque, NM 87131, USA}
\affiliation{Center for Computing Research (CCR),
             Quantum Algorithms and Applications Collaboratory (QuAAC),\\
             Department of Quantum Computer Science,
             Sandia National Laboratories,
             Albuquerque, NM 87185, USA}
\email{adbacze@sandia.gov}

\begin{abstract}
We present an architecture for early fault-tolerant quantum computers based on the ``smallest interesting colour code'' (Earl Campbell, 2016).
It realizes a universal logical gate set consisting of single-qubit measurements and preparations in the $X$ and $Z$ bases, single-qubit Hadamard ($H$) gates, and three-qubit controlled-controlled-$Z$ ($\CCZ$) gates.
State teleportations between \squarecode\ (2D) and \cubecode\ (3D) error-detecting color codes allow one to make use of the respective transversal $H$ and $\CCZ$ gates that  these codes possess.
As such, minimizing the number of logical quantum teleportation operations, not the number of logical quantum non-Clifford gates, is the relevant optimization goal.  
To help hardware developers characterize this architecture, we also provide an experimental protocol tailored to testing logical quantum circuits expressed in it. 

\end{abstract}

\maketitle

\section{Introduction}
\label{sec:intro}

Increasingly sophisticated demonstrations of the encoding and fault-tolerant manipulation of information stored in quantum error correcting (QEC) codes~\cite{PhysRevX.11.041058,  acharya2024quantumerrorcorrectionsurface, Bluvstein_2023, Wang_2024, pogorelov2024experimentalfaulttolerantcodeswitching, ryan2024high, reichardt2024demonstration, reichardt2024logical,lacroix2024scaling,eickbusch2024demonstrating,rodriguez2024experimental,bluvstein2025architectural,dasu2025breaking,daguerre2025experimental} make it clear that explorations of the capabilities of early fault-tolerant quantum computers (EFTQCs) will soon be feasible for both users and developers.
A body of theoretical work has also grown around the design of algorithms tailored to the prospective capabilities of EFTQCs~\cite{Campbell_2021,Ni_2023,QCELS,Ding_2023,liang2023modeling,PhysRevA.110.042420,anand2024stabilizerconfigurationinteractionfinding, GSP_EE_EFTQC, PRXQuantum.5.020101}. 
The performance-limiting features of EFTQC architectures that are typically considered include limited access to ancilla qubits and non-zero logical error rates, but relatively little consideration has been given to other features like non-uniformity in the overheads of implementing various logical operations.
One goal of this article is to propose a simple and specific architecture to serve as a starting point for exploring and designing to the capabilities of a small FTQC.
We employ distance-two codes that have already been demonstrated on EFT technologies~\cite{Bluvstein_2023,Wang_2024}.

QEC codes store reliable encoded logical qubits in the state of many error-prone physical qubits, achieving an exponential suppression in logical error rates with only a polynomial overhead in redundancy~\cite{PhysRevA.52.R2493, Good_QECs, aharonov1997fault, knill1998resilient}.
Fault-tolerant quantum computation on logical qubits requires not only that information be stored in QEC codes, but also that the logic gates processing the encoded information be implemented in such a way that small errors introduced in implementing these logical operations do not propagate catastrophically~\cite{shor1996fault}.
A straightforward method to achieve this is through the use of \emph{transversal} logical gates, namely logical gates that can be realized by acting on the physical qubits in each QEC code block separately and independently.  Because transversal logical gates never mix information within a QEC code block, they can spread errors at most between the logical qubits on which they act; this makes them intrinsically fault-tolerant, because one inserts an active quantum error correction step between every logical gate in a standard fault-tolerant quantum computing design~\cite{Preskill:1998a, Gottesman:2024a}.

By the Eastin-Knill theorem, no single QEC code can realize a universal logical gate set soley with transversal logical gates~\cite{eastin2009restrictions}.
However, there are multiple ways to achieve universality despite this restriction.
For example, \emph{gate teleportation}~\cite{Gottesman_1999, Zhou_2000} employs so-called magic states~\cite{Bravyi:2005a} to implement gates that aren't available transversally.
While the cost of preparing magic states was long viewed as one of the biggest bottlenecks in this approach to universal FTQC~\cite{PhysRevA.86.032324, Beverland_2020}, the attendant overheads have come down significantly in recent years~\cite{gidney2024magic}.
Alternatives to gate teleportation like \emph{code switching}~\cite{Paetznick:2013a, Anderson:2014b, Bombin:2013a, Bombin:2014b} use two or more distinct QEC codes with complimentary transversal gate sets that are combined to achieve universality.
However, when these complimentary QEC codes arise as different ``gauge fixings'' of a subsystem QEC code~\cite{PhysRevA.91.032330}, the fault-tolerant transfer of encoded information between the codes is generically accomplished through a sequence of elaborate operations that can be even more costly than distilling magic states~\cite{PRXQuantum.2.020341}.
An appealing middle ground is \emph{transversal code switching}~\cite{2024efficientfaulttolerantcodeswitching}, which is code switching facilitated by transversal gates.  
Not all code pairs can achieve this, but when they can, encoded quantum states can be teleported between the two codes using only transversal operations and measurements at a relatively low overhead cost~\cite{Paetznick:2013a}.

In this article, we present an EFTQC architecture that utilizes transversal code switching between the \squarecode\ (2D) and \cubecode\ (3D) error-detecting color codes~\cite{Bombin:2006a, Bombin:2007b, 832blog}.
We show that it is possible to compile arbitrary circuits into a gate set consisting solely of logical controlled-controlled-$Z$ ($\CCZ$) and Hadamard ($H$) gates~\cite{mikeIke}, a natural universal gate set for these codes considered as a code-switchable pair.
This architecture is restricted to distance-two codes and thus we do not expect it to enable utility-scale quantum computation~\cite{proctor2025benchmarking}.
However, it motivates cost models according to which the performance of EFTQCs could be evaluated.

The transition from physical circuits to logical circuits will require reassessing how metrics like computational volume~\cite{PhysRevA.100.032328, Blume_Kohout_2020} should be defined.
Our architecture provides an explicit and simple example in which the total number of one-, two-, and three-qubit gates in a compiled logical circuit, often considered to be a performance-limiting factor when it becomes too large, does not correlate well with actual performance-limiting factors, such as the number of logical teleportations in the compiled logical circuit.
To address mismatches like this, we conclude by presenting metrics that provide relevant logical performance information, and we further present characterization protocols for them that could be experimentally implemented in this architecture and others like it.

The remainder of this article is organized as follows: in Section~\ref{sec:background}, we review the universality of the $\{\CCZ, H\}$ gate set and survey the logical gates available in the \squarecode\ and \cubecode\ codes by transversal operations and permutations.
In Section~\ref{sec:compiler}, we  briefly discuss the compilation of arbitrary algorithms in the $\{\CCZ, H\}$ gate set by first synthesizing them into quantum circuits in the $\{\textrm{Clifford}, \Lambda(S)\}$ gate set using Ref.~\cite{glaudell2021optimaltwoqubitcircuitsuniversal} and then exactly synthesizing the $\{\textrm{Clifford}, \Lambda(S)\}$ gate set into the $\{\CCZ, H\}$ gate set.   
In Section~\ref{sec:architecture}, we show how to combine the results of Sections~\ref{sec:background} and~\ref{sec:compiler} to form a universal transversal EFTQC architecture.
In Section~\ref{sec:performance}, we introduce protocols one can use to characterize the performance of a processor utilizing our architecture in EFTQC hardware.  
Finally, in Section~\ref{sec:summary_future_work}, we summarize our results and present directions for future research.

\section{Background}
\label{sec:background}

\subsection{Notation}
\label{sec:notation}

We use the word \emph{qubit} to refer to a two-state system with Hilbert space $\CC^2$. 
We represent the \emph{standard qubit basis} in Dirac bra-ket notation as $|0\> = (1\ 0)^T$ and $|1\> = (0\ 1)^T$, where $(...)^T$ denotes the transpose of a row vector $(...)$. 
In this basis, we represent the identity and Pauli matrices $\boldsymbol{\sigma} = \left( X, Y, Z \right)$ with the shorthand

\begin{align}
I := \begin{pmatrix} 1 & 0\\ 0 & 1 \end{pmatrix}, \qquad
X := \begin{pmatrix} 0 & 1\\ 1 & 0 \end{pmatrix}, \qquad
Y := \begin{pmatrix} 0 & -i\\ i & 0 \end{pmatrix}, \qquad
Z := \begin{pmatrix} 1 & 0\\ 0 & -1 \end{pmatrix}.
\end{align}
Also in this basis, the standard matrices for the Hadamard, phase and $\pi/8$
gates take the form
\begin{align}
H &:=
   \frac{1}{\sqrt{2}} \begin{pmatrix} 1 & 1\\ 1 & -1 \end{pmatrix}
   &
S &:=
   \begin{pmatrix} 1 & 0\\ 0 & i \end{pmatrix}
   &
T &:=
   \begin{pmatrix} 1 & 0\\ 0 & e^{i\pi/4} \end{pmatrix}.
\end{align}

A projective (von Neumann) measurement of the Pauli observable $P \in \{X, Y, Z\}$ is denoted by $M_P$ and is represented by the pair orthogonal projectors $(I \pm P)/2$ with outcomes $\pm 1$.

Because quantum states are represented by rays in Hilbert space, we adopt the usual convention of representing states by their unit-norm representatives and ignoring any overall phase $e^{i\theta}$. 
Concomitantly, we ignore any overall phase in the action of any coherent gate, each of which is represented by unitary matrix so as to preserve the inner product between unit-norm representatives of rays.  This allows us to use 2D complex rotation-matrix representatives of the Pauli, $H$, $S$, and $T$ gates in $SU(2)$, the group of unit-determinant $2 \times 2$ unitary matrices.  A general 3D rotation about the $\mathbf{\hat{n}}$ axis by a counterclockwise angle $\theta$ when viewed from above is represented (nonuniquely) in $SU(2)$ by the following matrix:
\begin{align}
  R_{\mathbf{\hat{n}}}(\theta)
  &= \exp(-i\theta\, \mathbf{\hat{n}}\cdot \boldsymbol{\sigma}/2)
  = I \cos \frac{\theta}{2} - \mathbf{\hat{n}}\cdot\boldsymbol{\sigma} \sin\frac{\theta}{2}.
\end{align}
We use the shorthands $R_x$, $R_y$, and $R_z$ for rotations about the $x$, $y$, and $z$ axes.
This representation is why the $T$ gate is also called the $\pi/8$ gate; in this representation, the aforementioned matrices are:
\begin{align}
\tilde{I} = \begin{pmatrix} 1 & 0\\ 0 & 1 \end{pmatrix}, \qquad
\tilde{X} = \begin{pmatrix} 0 & i\\ i & 0 \end{pmatrix}, \qquad
\tilde{Y} = \begin{pmatrix} 0 & 1\\ -1 & 0 \end{pmatrix}, \qquad
\tilde{Z} = \begin{pmatrix} i & 0\\ 0 & -i \end{pmatrix}, \\[1.5ex]
\tilde{H} = \frac{1}{\sqrt{2}} \begin{pmatrix} i & i\\ i & -i \end{pmatrix}
\qquad
\tilde{S} = \begin{pmatrix} e^{-i\pi/4} & 0\\ 0 & e^{i\pi/4} \end{pmatrix}
\qquad
\tilde{T} = \begin{pmatrix} e^{-i\pi/8} & 0\\ 0 & e^{i\pi/8} \end{pmatrix}.
\end{align}

A (quantum-)controlled-$U$ gate, denoted by $\Lambda(U)$ is defined as
\begin{align}
\Lambda(U)|c\>|\psi\> :=
   \begin{cases}
      |0\>|\psi\>  & \text{if } c=0 \\
      |1\>U|\psi\>  & \text{if } c=1.
   \end{cases}
\end{align}
This definition applies even if the control space is one-dimensional (\ie, a phase).
In the language of controlled gates, the $Z$, $S$, and $T$ gates can be expressed as
\begin{align}
Z = \Lambda(-1), \qquad
S = \Lambda(i), \qquad
T = \Lambda(e^{i\pi/4}).
\label{eq:Lambda-ZST}
\end{align}
Some controlled two-qubit gates that we consider in this article include the controlled-NOT, controlled-phase, and controlled-$S$ gates:
\begin{align}
\textit{CX} := \textit{CNOT\/} := \Lambda(X), \qquad
\textit{CPHASE\/} := \Lambda(Z), \qquad
\textit{CS\/} := \Lambda(S).
\end{align}
We also consider the three-qubit Toffoli and controlled-controlled-$Z$ gates, where control on $k$ qubits is indicated by the notation $\Lambda^k$:
\begin{align}
\textit{TOF\/} = \Lambda^2(X) := \Lambda(\Lambda(X)), \qquad
\CCZ = \Lambda^2(Z) := \Lambda(\Lambda(Z)).
\end{align}

When it could be ambiguous as to which qubits are control qubits and which qubits are target qubits in a multi-qubit environment, we use subscript indices to clarify.  
For example, a Toffoli gate controlled by qubits $3$ and $5$ that acts on qubit $7$ as a target could be denoted as $\Lambda_3(\Lambda_5(X_7))$ or simply $\Lambda_{[3, 5]}(X_7)$.

The set of $n$-fold tensor products of Pauli matrices multiplied by phases from the set $\{\pm 1, \pm i\}$ form a group we denote by $\calP_n$.
This group is generated by the imaginary number $i$ and the Pauli $X$ and $Z$ operators on each qubit as
\begin{align}
\calP_n &= \<i\> \times \<X_1, Z_1, \ldots X_n, Z_n\>,
\end{align}
where $X_k$ and $Z_k$ denote operators that act as the Pauli $X$ and $Z$ operators on qubit $k$ and as the identity operator on all other $n-1$ qubits.  When it is clear from context, we use string concatenation to represent the tensor product of operators, such as $IX$ for $I \otimes X$, or equivalently $X_2$, in $\calP_2$.

The Clifford group $\calC_n$ is the group of matrices in $SU(2^n)$ that conjugate elements of the Pauli group to elements of the Pauli group:
\begin{align}
\calC_n &:= \{U \in SU(2^n)\ |\ \forall\,P \in \calP_n, UPU^{-1} \in \calP_n\}.
\end{align}
This group is a finite group, generated by, say, the $H$ and $S$ gates on arbitrary qubits and the $\CNOT$ gate between arbitrary pairs of qubits:
\begin{align}
\calC_n &= \<H, S, \CNOT \>.
\end{align}

We adopt the notation for quantum circuits from the textbook by Nielsen and Chuang~\cite{mikeIke}, in which time flows from left to right, qubits are ordered vertically, and single- and double-wires indicate quantum and classical bit worldlines, respectively.
We adopt the notation for QEC codes from the textbook by Gottesman~\cite{Gottesman:2024a}, in which $[\![n, k, d]\!]$ denotes a code encoding $k$ logical qubits into $n$ physical qubits with code distance $d$.  We use overbars to indicate gates on logical qubits, such as $\overline{X}$, $\overline{Y}$ and $\overline{Z}$.  
We omit a comprehensive review of the stabilizer formalism for QEC codes here, and refer the reader to Gottesman's textbook for details.

We sometimes describe preparation and measurements of qubits as ``gates,'' even though they are not coherent operations, and we often explicitly include them in the list of operations required to establish a universal gate set, namely one with which quantum circuits can approximate any solution to the \Schroedinger\ equation arbitrarily well.

\subsection{Universality of the \texorpdfstring{$\{\CCZ, H, M_Z, |0\>\}$}{\{CCZ, H, MZ, |0>\}} gate
basis via a phase-reference qubit}
\label{sec:ccz-h-universality}

A classic result in quantum circuit theory is that the gate set consisting of measurements and preparations of qubits in the $Z$ basis, along with the coherent Toffoli and Hadamard gates, is universal for quantum computation~\cite{Shi:2003a, aharonov2003simple}.

Because these matrices do not contain imaginary numbers, while the \Schroedinger\ equation does, it may seem that it would be impossible for such a gate set to be universal.
However, the universality becomes apparent if we encode the imaginary component of all numbers into a global ``phase-reference'' qubit $p$.
With that phase-reference qubit, given a decomposition of a quantum state into real and imaginary components as
\begin{align}
|\psi\> &=
  \textrm{Re}\,|\psi\> +
  i\,\textrm{Im}\,|\psi\>,
\end{align}
the phase-reference encoding of the state is 
\begin{align}
\underline{|\psi\>} &:=
  |0\>_p \otimes
  \textrm{Re}\,|\psi\>
  +
  |1\>_p \otimes
  \textrm{Im}\,|\psi\>.
\end{align}

In this representation, a (complex) unitary matrix $U$ applied to $|\psi\>$ acts as the (real) unitary matrix $\underline{U}$ on $\underline{|\psi\>}$ does: 
\begin{align}
\nonumber
  \underline{U}
  &=
  \phantom{+}\
    \left( |0\>\<0| + |1\>\<1| \right)_p \otimes
    \textrm{Re}\,U 
   +
    \left(-|0\>\<1| + |1\>\<0| \right)_p \otimes
    \textrm{Im}\,U\\
  &=
  \phantom{+}\
    I \otimes
    \textrm{Re}\,U
    -iY \otimes
    \textrm{Im}\,U. \label{eq:pre}
\end{align}
%
This phase-reference encoding is a faithful map of unitary matrices in $SU(2^n)$ to unitary matrices in $SU(2^{n+1})$, and is therefore always a valid quantum gate~\cite{kissinger2024catalysingcompletenessuniversality}.
This simulates the algebra of complex numbers as follows:
\begin{align}
  |\psi'\> &= U|\psi\> \nonumber \\
 &=
  (\textrm{Re}\, U + i\,\textrm{Im}\, U)
  (\textrm{Re}\,|\psi\> + i\,\textrm{Im}\,|\psi\>) \nonumber \\
  &= 
    (\textrm{Re}\, U)
      (\textrm{Re}\,|\psi\> + i\,\textrm{Im}\,|\psi\>)
  -
    (\textrm{Im}\, U)
      (\textrm{Im}\,|\psi\> - i\,\textrm{Re}\,|\psi\>).
\end{align}

Although the classic result uses the Toffoli gate, it manifestly also applies to the $\CCZ$ gate because
\begin{align}
  \CCZ
  &= \Lambda^2(Z) \nonumber \\
  &= \Lambda^2(HXH) \nonumber \\
  &= H_3\, \Lambda^2(X)\, H_3 \nonumber \\
  &= H_3\,\TOF\,H_3.
\end{align}

As a quantum circuit, this relation is expressed as

\begin{figure}[!ht]
    \centering
    \includegraphics{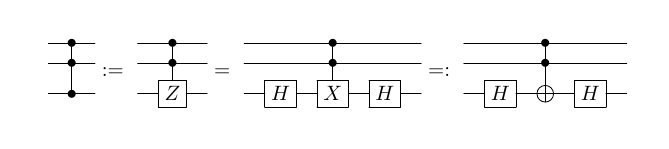}
    \caption{Conjugating the target of the Toffoli gate with Hadamard gates yields the $\Lambda^2(Z)$ gate}
\end{figure}

One can prove the universality of the $\{\CCZ, H\}$ gate set in many ways. 
For example, it is straightforward to show that one can simulate the Kitaev universal $\{\Lambda(S), H\}$ gate set~\cite{Kitaev_1997, aharonov2003simple, kissinger2024catalysingcompletenessuniversality}, exactly with it.
Using Equation~(\ref{eq:pre}), it can be seen that to effect $\Lambda(S)$, $-iY$ should be applied to the phase-reference when both control and target take the value $|1\rangle$.
This can be done with two $\CCZ$ and $H$ gates, using the circuit depicted in Figure~\ref{fig:CS}.

\begin{align}
\Lambda(S) =
\begin{pmatrix}
 1 &   &   &   \\
  & 1 &   &   \\
  &   & 1 &   \\
  &   &   & i \\
\end{pmatrix}
&\overset{\textrm{Equation~(\ref{eq:pre})}}{\longmapsto}
\begin{pmatrix}
 1 &   &   &   &   &   &   &   \\
  & 1 &   &   &   &   &   &   \\
  &   & 1 &   &   &   &   &   \\
  &   &   & 0 &   &   &   & -1\\
  &   &   &   & 1 &   &   &   \\
  &   &   &   &   & 1 &   &   \\
  &   &   &   &   &   & 1 &   \\
  &   &   & 1 &   &   &   & 0 \\
\end{pmatrix}
=
\Lambda_3(\Lambda_2(-iY_1))
\end{align}
\begin{figure}[H]
\begin{align}
\nonumber
\raisebox{-1ex}{%
\Qcircuit @C=1em @R=1em {
 & \ctrl{1} & \qw \\
 & \gate{\underline{S}} & \qw
} 
} 
\quad
\raisebox{-4ex}{=}
\qquad
\Qcircuit @C=1em @R=1em {
 \lstick{p} & \gate{-iY} & \qw \\
 & \ctrl{-1} & \qw \\
 & \ctrl{-1} & \qw
} 
\quad
\raisebox{-4ex}{=}
\qquad
\Qcircuit @C=1em @R=1em {
 \lstick{p} & \control \qw  & \gate{H} & \control \qw  & \gate{H} & \qw \\
            & \control \qw  & \qw      & \control \qw  & \qw      & \qw \\
            & \ctrl{-2} & \qw      & \ctrl{-2} & \qw      & \qw
} 
\end{align}
    \caption{\label{fig:CS} $\Lambda(\underline{S})$ gate decomposed in the phase reference encoding using $\{\CCZ, H\}$ as prescribed by Equation~(\ref{eq:pre}). The phase-reference qubit is labeled by $p$.}
\end{figure}

\subsection{Logical gates in the \texorpdfstring{\squarecode}{[[4, 2, 2]]} and
\texorpdfstring{\cubecode}{[[8, 3, 2]]} color codes}
\label{sec:422-832-logical-gates}

\begin{figure}[!ht]
    \centering
    \includegraphics[width=0.35\textwidth]{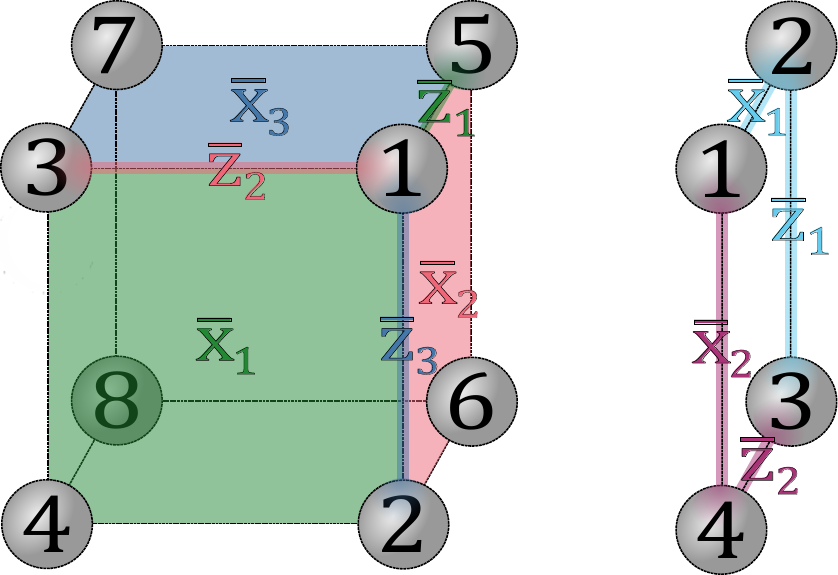}
    \caption{Encoded logical operators of the \cubecode\ code (left) and the \squarecode\ code (right). Circles depict physical qubits and the colored planes/edges depict logical operators.}
    \label{fig:cube-square}
\end{figure}

The \squarecode\ and \cubecode\ codes are small examples of 2D~\cite{Bombin:2006a} and 3D~\cite{Bombin:2007b} color codes, respectively, with the latter having been dubbed by Earl Campbell as ``the smallest interesting (3D) colour code''~\cite{832blog}.  
There are other quantum codes with these parameters; for example, the 2D surface code on the cube is an \cubecode\ code~\cite{Landahl:2020a}.  When we refer to ``the'' \squarecode\ and ``the'' \cubecode\ codes moving forward, we will mean the 2D and 3D color codes with these parameters.
The \squarecode\ and \cubecode\ color codes are naturally represented by a square and a cube, respectively, which are special cases of the more general ``colex'' representations of color codes~\cite{Bombin:2007a}.
We depict these codes graphically in Figure~\ref{fig:cube-square}, and algebraically in Table~\ref{tab:codes}.
We will make frequent use of these graphical representations throughout this article.
As shown in Figure~\ref{fig:cube-square}, we display physical qubits as gray circles and encoded logical operators as colored edges and faces.  These codes have already been put to use in special-purpose EFTQC hardware, including demonstrations of preparations of logical graph states and one-bit quantum adders~\cite{reichardt2024logical, Wang_2024}.

\begin{table}[!ht]
 \centering
 \setlength{\extrarowheight}{5pt} 
    \begin{tabular}{|c|c|c|c|}
        \hline
        & \raisebox{0.235em}{\cubecode\ }  & \raisebox{0.235em}{\squarecode\ } \\
        \hline
        \parbox[t]{3mm}{\multirow{5}{*}{\rotatebox[origin=c]{90}{Stabilizers}}} & $S_1 = 
         \sx{Z}\sx{Z}\sx{Z}\sx{Z}\sx{I}\sx{I}\sx{I}\sx{I}$ & $S_1 = \sx{X}\sx{X}\sx{X}\sx{X}$
         \\
         & $S_2 = \sx{Z}\sx{Z}\sx{I}\sx{I}\sx{Z}\sx{Z}\sx{I}\sx{I}$ & $S_2 = \sx{Z}\sx{Z}\sx{Z}\sx{Z}$ \\ 
         & $S_3 = \sx{Z}\sx{I}\sx{Z}\sx{I}\sx{Z}\sx{I}\sx{Z}\sx{I}$ & \\
         & $S_4 = \sx{Z}\sx{Z}\sx{Z}\sx{Z}\sx{Z}\sx{Z}\sx{Z}\sx{Z}$ & \\
         & $S_5 = \sx{X}\sx{X}\sx{X}\sx{X}\sx{X}\sx{X}\sx{X}\sx{X}$ &  \\
         \hline 
         \parbox[t]{3mm}{\multirow{6}{*}{\rotatebox[origin=c]{90}{Logical Operators}}}
         & \textcolor{mygreen}{$\overline{X}_1 = \sx{X}\sx{X}\sx{X}\sx{X}\sx{I}\sx{I}\sx{I}\sx{I}$} & \textcolor{mycyan}{$\overline{X}_1 = \sx{X}\sx{X}\sx{I}\sx{I}$} \\ 
         & \textcolor{mygreen}{$\overline{Z}_1 = \sx{Z}\sx{I}\sx{I}\sx{I}\sx{Z}\sx{I}\sx{I}\sx{I}$} & \textcolor{mycyan}{$\overline{Z}_1 = \sx{I}\sx{Z}\sx{Z}\sx{I}$} \\
         & \textcolor{myred}{$\overline{X}_2 = \sx{X}\sx{X}\sx{I}\sx{I}\sx{X}\sx{X}\sx{I}\sx{I}$} & \textcolor{mypurple}{$\overline{X}_2 = \sx{X}\sx{I}\sx{I}\sx{X}$} \\
         & \textcolor{myred}{$\overline{Z}_2 = \sx{Z}\sx{I}\sx{Z}\sx{I}\sx{I}\sx{I}\sx{I}\sx{I}$} & \textcolor{mypurple}{$\overline{Z}_2 = \sx{I}\sx{I}\sx{Z}\sx{Z}$} \\
         & \textcolor{myblue}{$\overline{X}_3 = \sx{X}\sx{I}\sx{X}\sx{I}\sx{X}\sx{I}\sx{X}\sx{I}$} & \\
         & \textcolor{myblue}{$\overline{Z}_3 = \sx{Z}\sx{Z}\sx{I}\sx{I}\sx{I}\sx{I}\sx{I}\sx{I}$} & \\
 \hline
 \end{tabular}
 \caption{Algebraic representation of the stabilizer generators and logical operators of the \cubecode\ (left) and \squarecode\ (right) codes. The color of the text depicting the logical operators is chosen to match the corresponding colored edges/faces in the graphical representation displayed in Figure~\ref{fig:cube-square}.}
 \label{tab:codes}
 \end{table}

Taken collectively as a pair, these two codes can realize a fault-tolerant universal gate-set transversally.
Furthermore, a transversal logical $\Lambda({X})^{\otimes 2}$ gate exists between any two logical qubits in the \cubecode\ code block as the controls and the two logical qubits in the \squarecode\ code block as the targets~\cite{Wang_2024}.
This transversal logical $\Lambda({X})^{\otimes 2}$ gate allows one to teleport quantum states between the codes, opening up transversal code switching that facilitates the universal gate set they can jointly realize.  A related code-switching operation has been demonstrated in an EFTQC platform using $[\![15, 1, 3]\!]$ and $[\![7, 1, 3]\!]$ codes as a pair to realize the universal $\{\textrm{Clifford}, T\}$ gate set transversally~\cite{bluvstein2025architectural}.
In the next subsections, we describe how our universal gate set is realized at the level of physical qubits.

\subsubsection{Transversal and Passive Permutation Gates}
\label{sec:transversal-gates}

\begin{figure}[!ht]
    \centering
    \begin{align*}
      \includegraphics[width=0.15\textwidth]{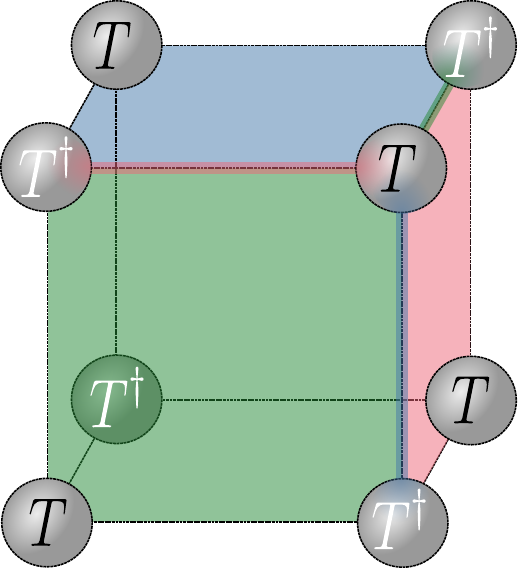}
      \qquad \raisebox{8ex}{=} \qquad 
      \raisebox{8ex}{$\overline{\CCZ}\left(\rule{0em}{10ex}\right.$
      \raisebox{-8ex}{\includegraphics[width=0.15\textwidth]{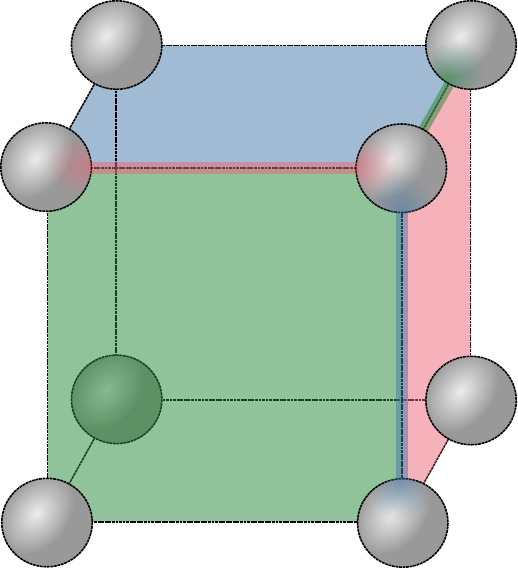}}
      $\left.\rule{0em}{10ex}\right)$}
    \end{align*}
    \caption{Logical \CCZ~gate between the three logical qubits encoded in a single \cubecode\ code block consisting of eight physical $T/T^{\dag}$ gates. $T$ gates and $T^{\dag}$ gates are applied in an alternating fashion to all physical qubits on the block.}
    \label{fig:log_ccz}
\end{figure}

One can fault-tolerantly implement the logical $\CCZ$ gate  on the \cubecode\ code by applying $T$ and $T^{\dag}$ gates in an alternating pattern to the physical qubits~\cite{832blog}, as depicted in Figure~\ref{fig:log_ccz}.

\begin{figure}[!ht]
    \centering
    \begin{align*}
      &\includegraphics[width=0.15\textwidth]{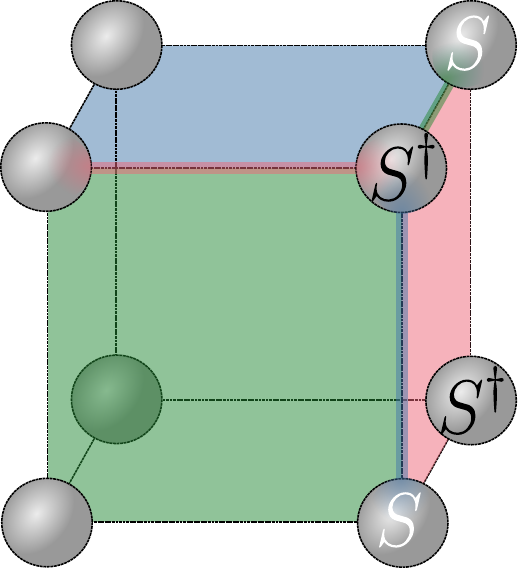}
      \qquad \raisebox{8ex}{$\cong$} \qquad 
      {\includegraphics[width=0.15\textwidth]{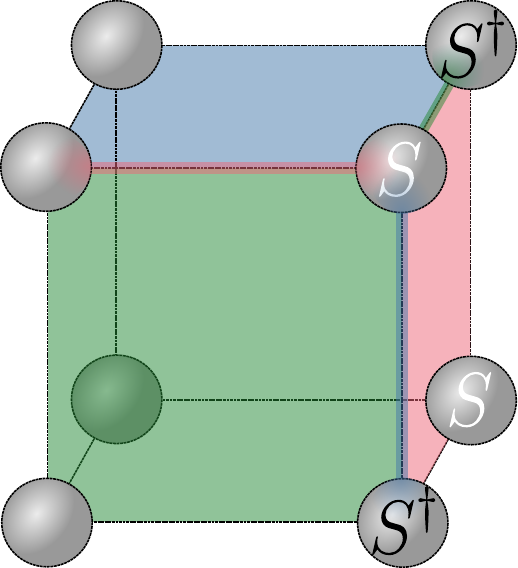}}
      \qquad \raisebox{8ex}{$\cong$} \qquad \\
      &{\includegraphics[width=0.15\textwidth]{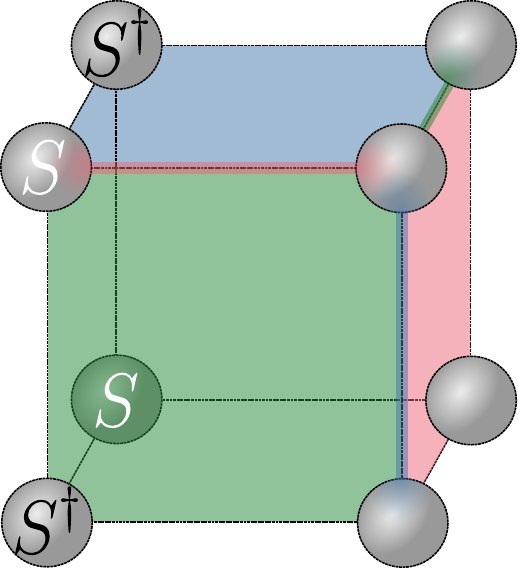}}
      \qquad \raisebox{8ex}{$\cong$} \qquad 
      {\includegraphics[width=0.15\textwidth]{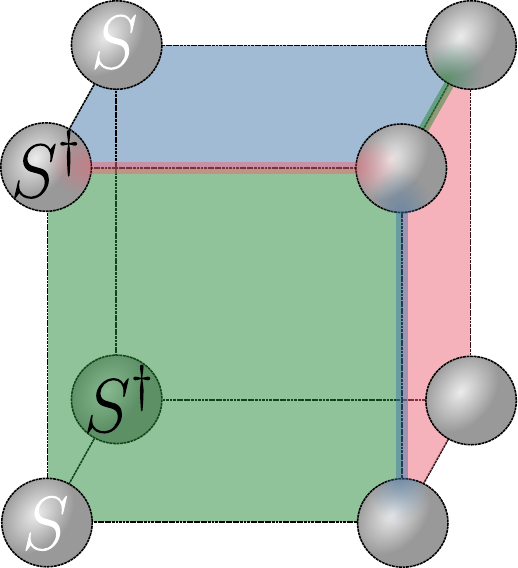}}
      \qquad \raisebox{8ex}{$\cong$} \qquad 
      \raisebox{8ex}{$\overline{\textit{CZ}}_{13}\left(\rule{0em}{10ex}\right.$
      \raisebox{-8ex}{\includegraphics[width=0.15\textwidth]{figures/cube_no_labels.pdf}}
      $\left.\rule{0em}{10ex}\right)$}
    \end{align*}
    \caption{Logical $\Lambda(Z)$ gate between two logical qubits encoded in a single \cubecode\ code block consisting of four physical $S/S^{\dag}$ gates. Here $\cong$ denotes equivalence under stabilizer element multiplication. The $S$ and $S^{\dag}$ gates are applied in alternating fashion around the face containing the full support of the logical $Z$ operators involved. This leads to four constructions equivalent under stabilizer element multiplication. The $\Lambda(Z)$ gate corresponding to the choice of physical qubits in the figure is between logical qubits $1$ (green) and $3$ (blue).}
    \label{fig:log_cz}
\end{figure}

Likewise, one can fault-tolerantly implement the logical $\Lambda(Z)$ gate between any two logical qubits in the same \cubecode\ code block by applying $S$ and $S^{\dag}$ gates in an alternating pattern to the four physical qubits on the face of cube containing the full support of the logical $Z$ operators of the qubits involved in the $\Lambda(Z)$ gate~\cite{Chen_2022}. 
Figure~\ref{fig:log_cz} depicts an example of such $\Lambda(Z)$ gate.  
These four qubits are equivalently described as being on the face containing the support of the logical $X$ operator on the logical qubit \emph{not} participating in the $\Lambda(Z)$ gate.  
Using the stabilizer generator $S_2$ from Table~\ref{tab:codes}, one can also construct an equivalent version with the $S$ and $S^{\dagger}$ gates reversed, because $ZS = S^\dagger$.
Similarly, through multiplication by the stabilizer generators, the parallel faces and edges of the cube also contain the logical operators, so there are two more constructions of the $\Lambda(Z)$ gate that one can realize by applying the physical $S/S^{\dag}$ gates to the parallel face of the code block.  Figure~\ref{fig:log_cz} captures all four of these possibilities.

\begin{figure}[!ht]
    \centering
    \begin{align*}
      \includegraphics[width=0.4\textwidth]{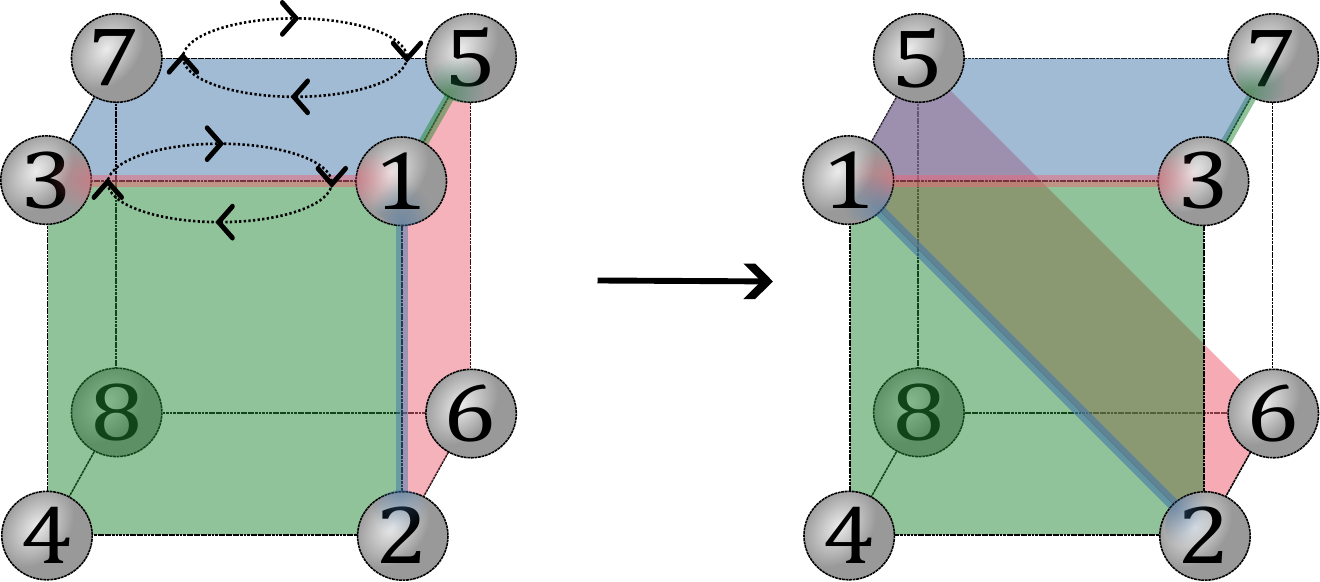}
      \qquad \raisebox{8ex}{$=$} \qquad 
      \raisebox{8ex}{$\overline{\CNOT}_{23}\left(\rule{0em}{10ex}\right.$
      \raisebox{-8ex}{\includegraphics[width=0.15\textwidth]{figures/cube_no_labels.pdf}}
      $\left.\rule{0em}{10ex}\right)$}
    \end{align*}
    \caption{Logical $\Lambda(X)$ gate between two logical qubits encoded in a \cubecode\ code block realized by permuting/relabeling physical qubits. In this figure, logical qubit 2 (red) is the control qubit and logical qubit 3 (blue) is the target qubit.}
    \label{fig:log_cnot}
\end{figure}

In principle, one can fault-tolerantly implement the logical $\Lambda(X)$ gate between the two logical qubits in a single \cubecode\ code block by conjugating the target logical qubit of the logical $\Lambda(Z)$ gate by the logical $H$ gate.  However, we have not yet described how to implement the logical $H$ gate fault-tolerantly on a logical qubit in the \cubecode\ code.  We can sidestep this by using a different approach that only requires the physical qubits to be permuted/relabeled in a ``passive'' way, in other words, in a way that doesn't require any active gates at all~\cite{Wang_2024}.
Figure~\ref{fig:log_cnot} depicts an example how this approach works.  
As is well-known in stabilizer coding theory ~\cite{Gottesman:2024a}, the $\Lambda(X)$ gate causes the control and target Pauli-group generators to propagate as follows in the Heisenberg picture: $XI \leftrightarrow XX$, $IX \leftrightarrow IX$, $ZI \leftrightarrow ZI$, $IZ \leftrightarrow ZZ$.  
By permuting the physical qubits in the \cubecode\ code block as depicted in Figure~\ref{fig:log_cnot}, one can generate this action: the edge containing $\overline{Z}_{\textrm{target}}$ gains a component along $\overline{Z}_{\textrm{control}}$ and the plane containing $\overline{X}_{\textrm{control}}$ gains a component along $\overline{X}_{\textrm{target}}$.

\begin{figure}[!ht]
    \centering
    \begin{align*}
      &\mbox{\includegraphics[width=0.4\textwidth]{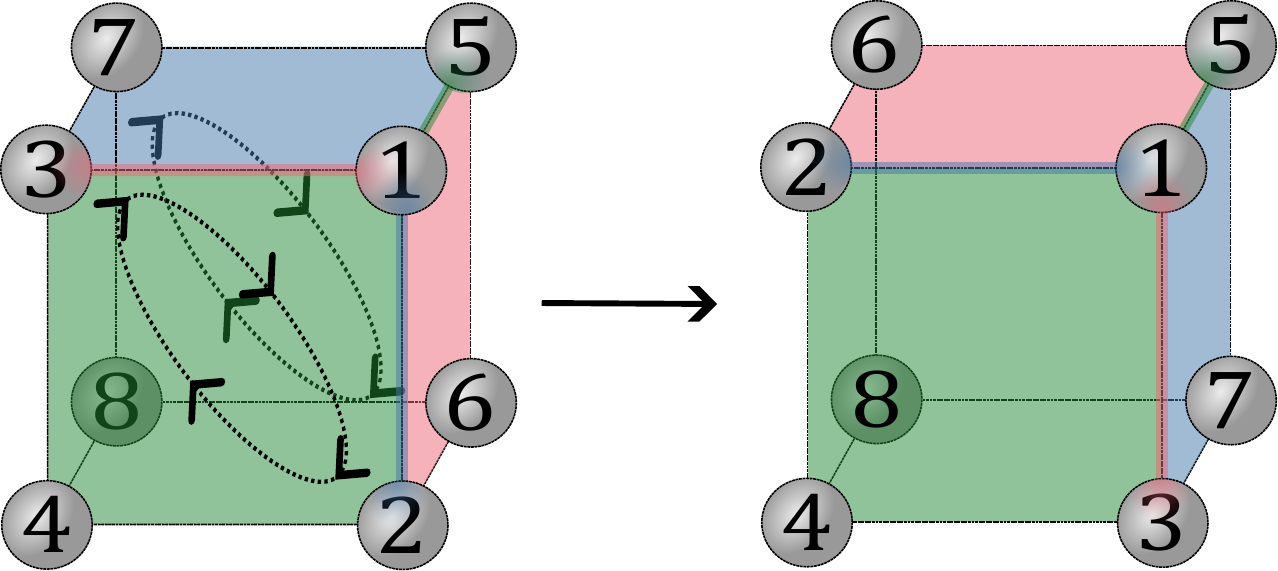}}
      \qquad \raisebox{8ex}{$\cong$} \qquad 
      \raisebox{-1.8em}{\includegraphics[width=0.5\textwidth]{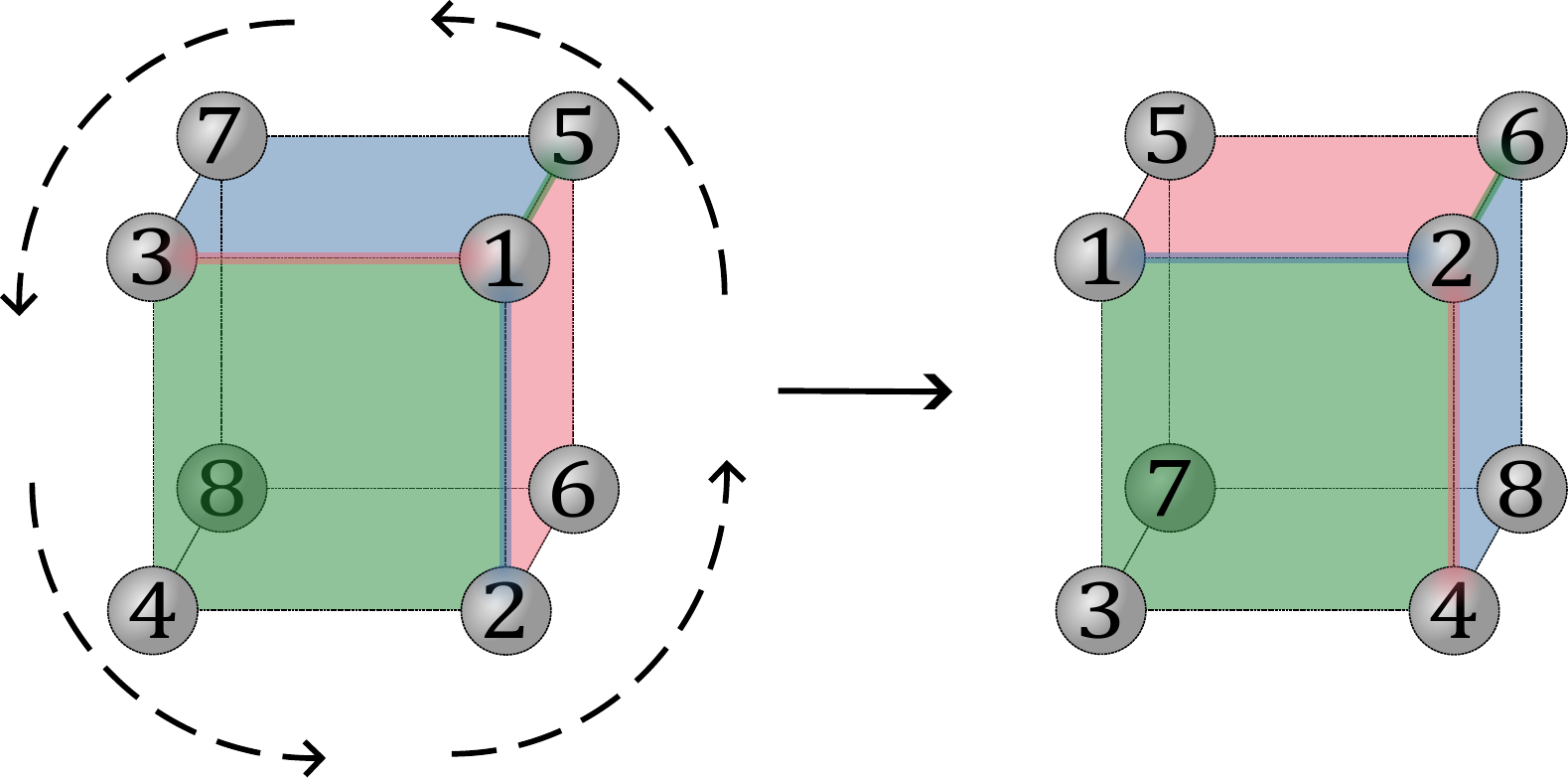}} \\
      & \raisebox{8ex}{$\cong$} \qquad 
      \raisebox{-1.8em}{\includegraphics[width=0.5\textwidth]{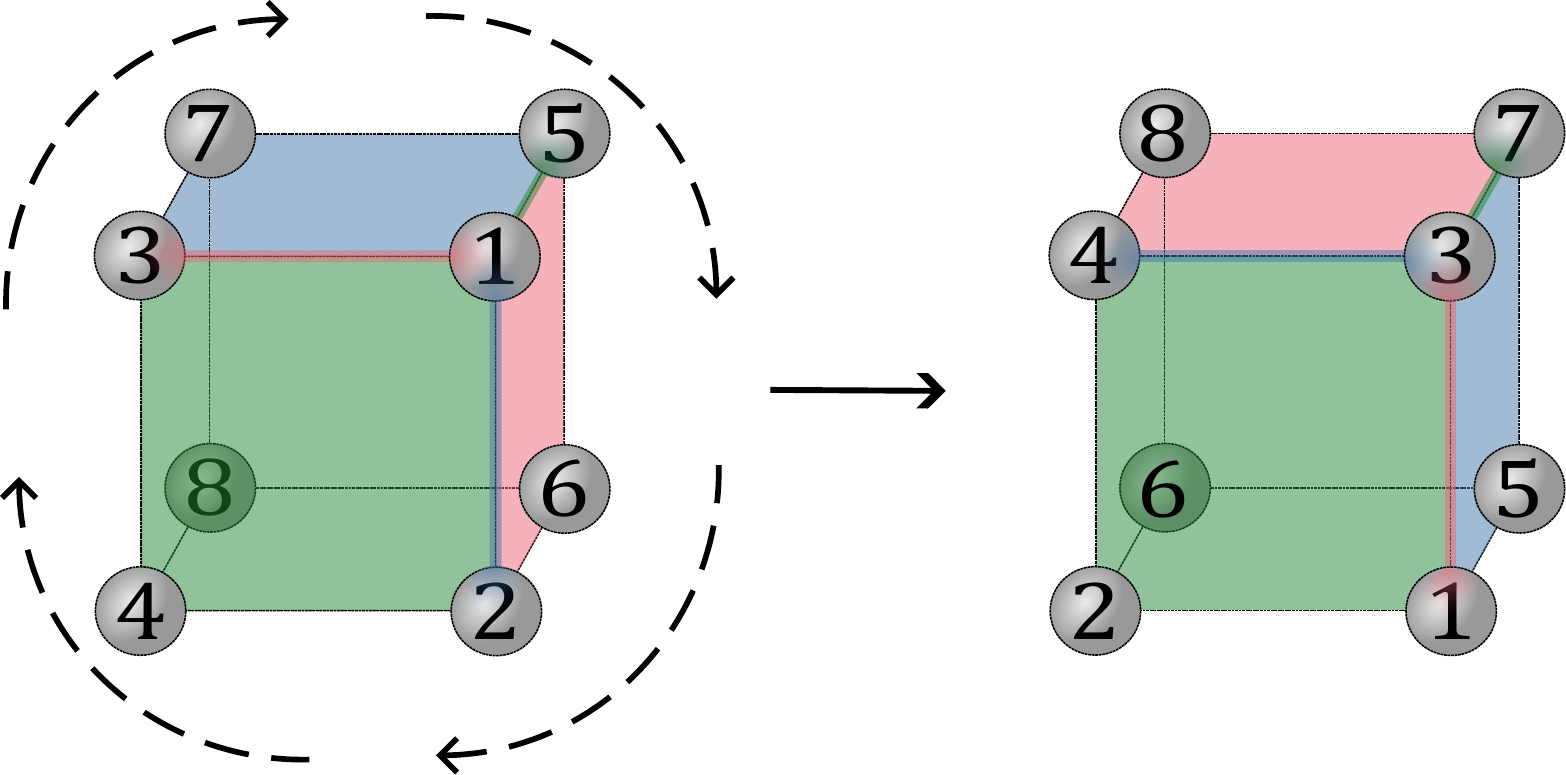}}
      \qquad \raisebox{8ex}{$\cong$} \qquad 
      \raisebox{8ex}{$\overline{\SWAP}_{23}\left(\rule{0em}{10ex}\right.$
      \raisebox{-8ex}{\includegraphics[width=0.15\textwidth]{figures/cube_no_labels.pdf}}
      $\left.\rule{0em}{10ex}\right)$}
    \end{align*}
    \caption{Logical $\SWAP$ gate between two logical qubits encoded in a \cubecode\ code block realized by permuting/relabeling physical qubits. In this figure, logical qubit 2 (red) and logical qubit 3 (blue) are swapped. This is enacted by permuting qubits along a diagonal plane which is equivalent, up to stabilizer multiplication, to rotating the cube by $\pm\, 90^{\circ}$. Here $\cong$ denotes equivalence up to stabilizer element multiplication.}
    \label{fig:832_SWAP}
\end{figure}

The $\SWAP$ gate can be formed via an alternating sequence of three $\CNOT$ gates; therefore the logical $\overline{\SWAP}$ gate can also be enacted by permuting the physical qubits.
As shown in Figure~\ref{fig:832_SWAP}, permuting qubits along a diagonal plane swaps the two logical qubits whose $\overline{X}$ and $\overline{Z}$ logical operators are both at half support along the diagonal plane.
This is equivalent, up to multiplication by stabilizer elements, to rotating the entire cube $\pm\, 90^{\circ}$ around the axis defined by the logical $\overline{Z}$ operator not involved in the $\SWAP$ gate.

\begin{figure}[!ht]
    \centering
    \begin{align*}
      \includegraphics[width=0.165\textwidth]{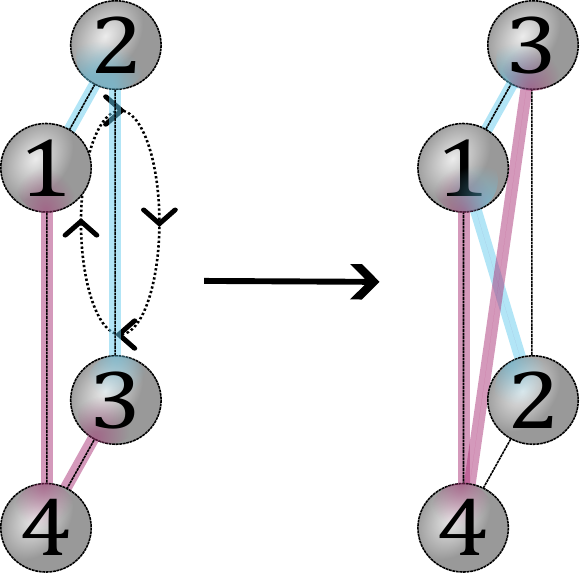}
      \qquad \raisebox{8ex}{$=$} \qquad 
      \raisebox{8ex}{$\overline{\CNOT}_{12}\left(\rule{0em}{10ex}\right.$
      \raisebox{-7.75ex}{\includegraphics[width=0.0465\textwidth]{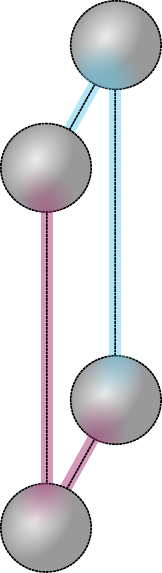}}
      $\left.\rule{0em}{10ex}\right)$}
    \end{align*}
    \caption{Logical $\Lambda(X)$ gate between two logical qubits encoded in a \squarecode\ code block realized by permuting/relabeling physical qubits. In this figure, logical qubit 1 (purple) is the control qubit and logical qubit 2 (cyan) is the target qubit.}
    \label{fig:log_cnot_422}
\end{figure}

\begin{figure}[!ht]
    \centering
    \begin{align*}
      \includegraphics[width=0.165\textwidth]{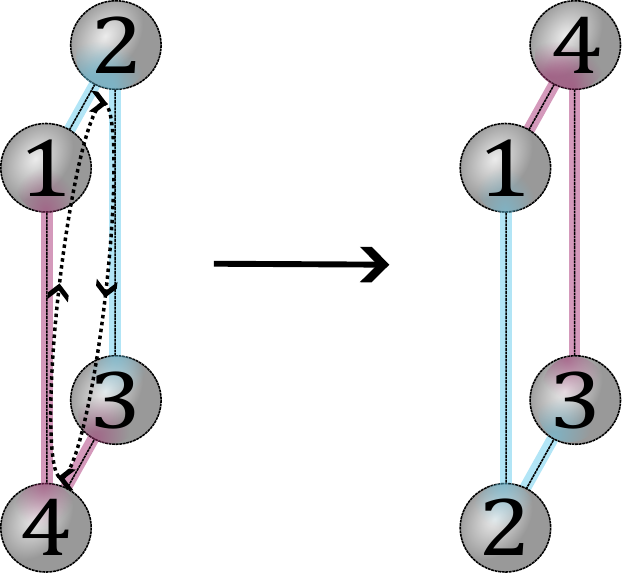}
      \qquad &\raisebox{8ex}{$\cong$} \qquad 
      \raisebox{-0.8em}{\includegraphics[width=0.195\textwidth]{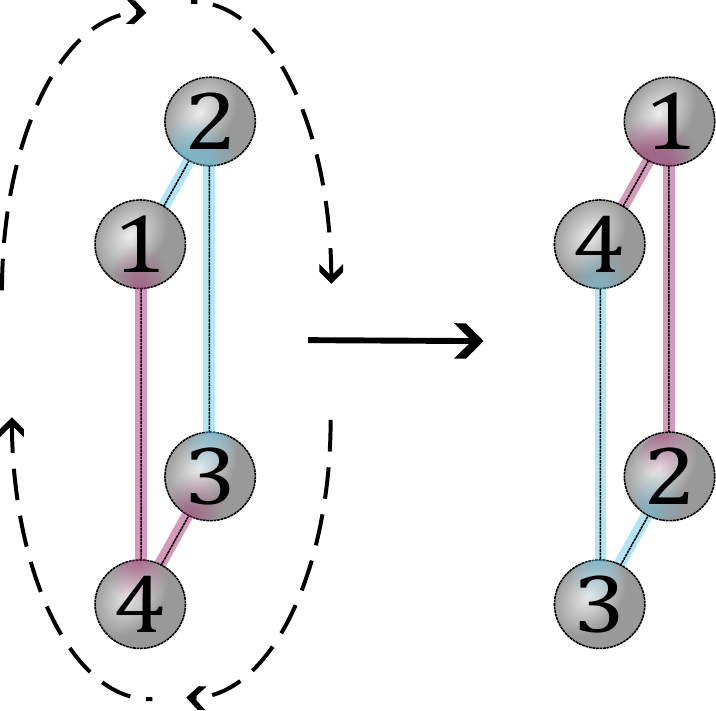}}
      \qquad \raisebox{8ex}{$\cong$} \qquad \\
       \raisebox{-0.8em}{\includegraphics[width=0.195\textwidth]{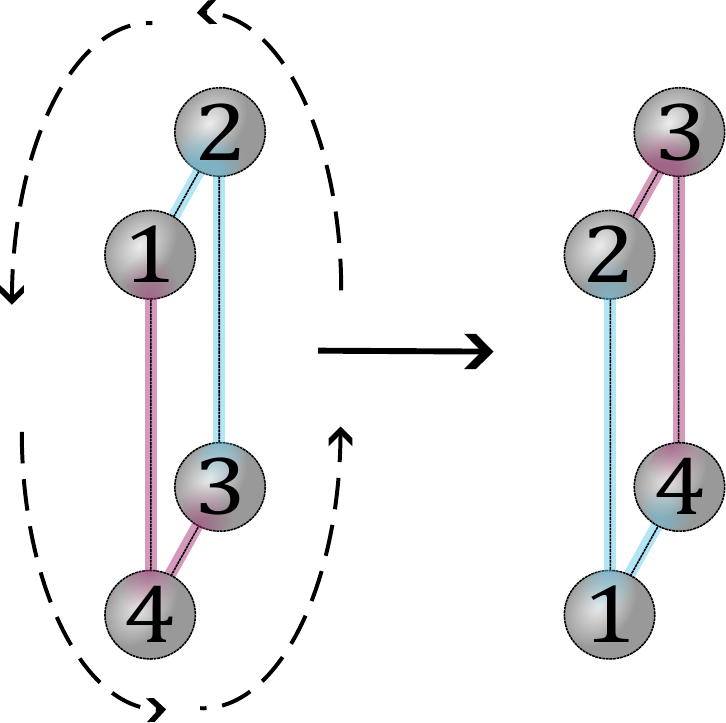}}
      \qquad &\raisebox{8ex}{$\cong$} \qquad 
      \raisebox{8ex}{$\overline{\SWAP}\left(\rule{0em}{10ex}\right.$
      \raisebox{-8ex}{\includegraphics[width=0.0465\textwidth]{figures/square_no_labels.pdf}}
      $\left.\rule{0em}{10ex}\right)$}
    \end{align*}
    \caption{Logical $\SWAP$ gate between the two logical qubits encoded in a \squarecode\ code block realized by permuting/relabeling physical qubits. This is enacted by permuting qubits along the diagonal which is equivalent, up to stabilizer element multiplication, to rotating the square by $\pm\, 90^{\circ}$. Here $\cong$ denotes equivalence up to stabilizer multiplication.}
    \label{fig:422_SWAP}
\end{figure}

Similarly, a $\Lambda(X)$ gate can be implemented in the \squarecode\ code block by a passive permutation. 
Figure~\ref{fig:log_cnot_422} depicts an example of such a gate.
Additionally, as shown in Figure~\ref{fig:422_SWAP}, permuting the diagonal physical qubits swaps the logical qubits, which is equivalent to rotating the code block by $\pm\, 90^{\circ}$, up to stabilizer multiplication.

The \cubecode\ code does not have a transversal logical $H$ gate, but the \squarecode\ code does.
Applying the physical $H$ gate to all qubits transversally in the \squarecode\ code block will swap the logical qubits and apply an $H$ gate to each of them.  In other words, the transversal $H$ gate implements the logical $(H \otimes H)\, \SWAP$ gate.
The $\SWAP$ gate can be undone using the aforementioned protocol shown in Figure~\ref{fig:422_SWAP}. 

\begin{figure}[!ht]
    \centering
    \begin{align*}
      {\includegraphics[width=0.0465\textwidth]{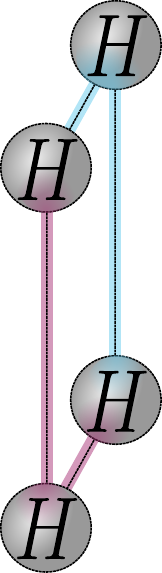}}
      \qquad \raisebox{8ex}{$=$} \qquad 
      \raisebox{8ex}{$(\overline{H}\otimes\overline{H})\overline{\SWAP}\left(\rule{0em}{10ex}\right.$
      \raisebox{-8ex}{\includegraphics[width=0.0465\textwidth]{figures/square_no_labels.pdf}}
      $\left.\rule{0em}{10ex}\right)$}
    \end{align*}
    \caption{Logical $(H \otimes H)\, \SWAP$ gate on the two logical qubits encoded in a single \squarecode\ code block.  Physical $H$ gates are applied to all qubits in the code block.}
    \label{fig:log_H}
\end{figure}

\begin{figure}[!ht]
    \begin{align*}
      \raisebox{-0.5em}{{\includegraphics[width=0.17\textwidth]{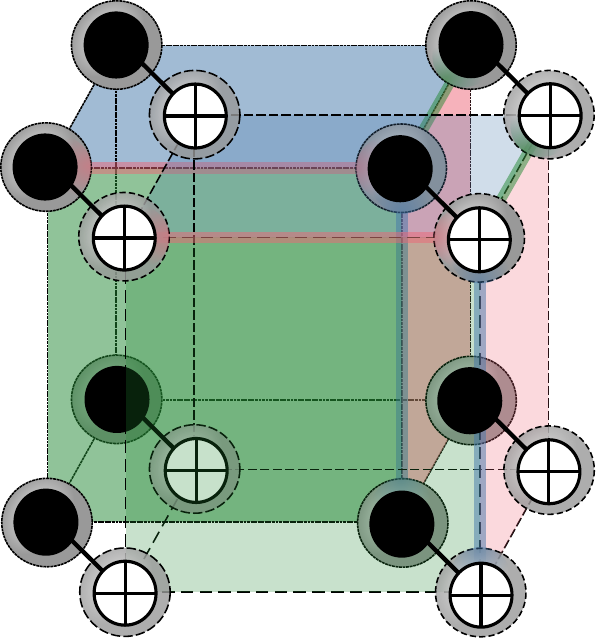}}}
      \qquad \raisebox{8ex}{$=$} \qquad 
      \raisebox{8ex}{$\Lambda(X)^{\otimes 3}\left(\rule{0em}{10ex}\right.$
      \raisebox{-8ex}{\includegraphics[width=0.3\textwidth]{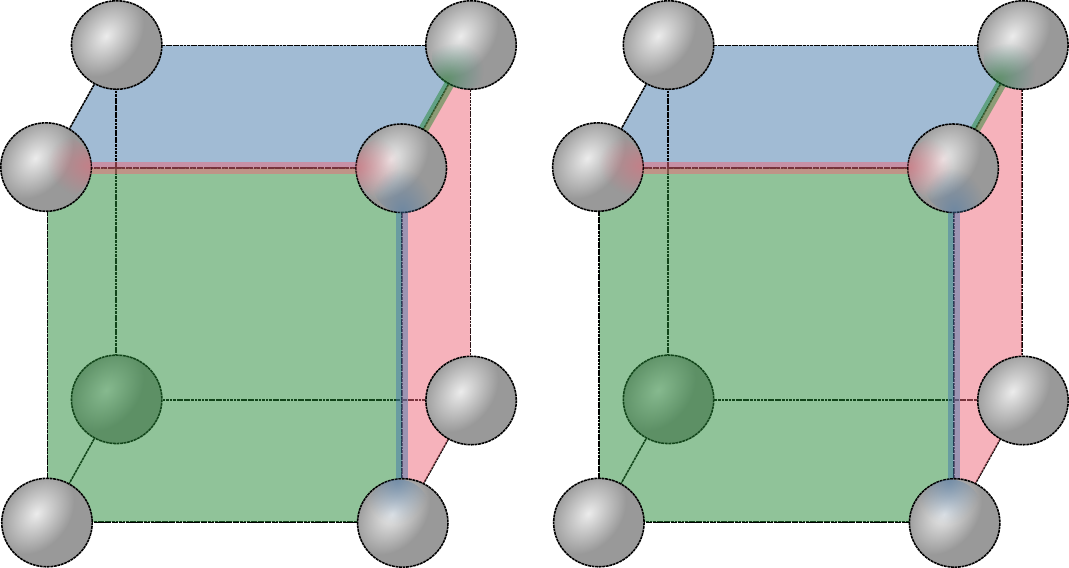}}
      $\left.\rule{0em}{10ex}\right)$}
    \end{align*}
    \caption{The logical $\Lambda(X)^{\otimes 3}$ gate between two \cubecode\ code blocks.  Relative to the choice of logical operators in Table~\ref{tab:codes}, this is facilitated by transversal $\Lambda(X)$ gates from the target logical block to the control target block.}
    \label{fig:triple_CNOT_dif_blocks}
\end{figure}

\begin{figure}[!ht]
    \centering
    \includegraphics{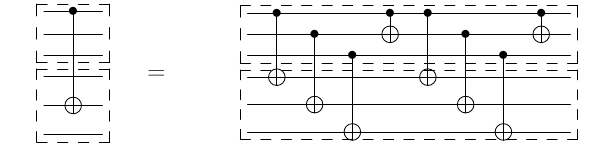}
    \caption{The logical $\Lambda(X)$ between the first logical qubit in one \cubecode\ code block and the second logical qubit in another. The dashed boxes depict the two separate sets of three logical qubits encoded in each \cubecode\ code block.}
    \label{fig:CNOT_dif_blocks_circ}
\end{figure}

A seminal result in QEC theory is that an $[\![n, k, d]\!]$ code that is a CSS code~\cite{Calderbank:1996a, Steane:1996b}, namely one whose stabilizer generators are each solely of Pauli $X$ or $Z$ type, such as the \squarecode\ and \cubecode\ codes listed in Table~\ref{tab:codes}, admits a transversal fault-tolerant logical $\Lambda(X)^{\otimes k}$ gate between all $k$ logical qubits in the pair of code blocks~\cite{mikeIke, Gottesman:2024a}.  Hence, one can implement the logical $\Lambda(X)^{\otimes 3}$ gate between two \cubecode\ code blocks transversally.  As shown in Ref.~\cite{hangleiter2024faulttolerant}, this can be combined with in-block $\Lambda(X)$ gates to enable a single logical $\Lambda(X)$ gate between an arbitrary single logical qubit in one  \cubecode\ code block and an arbitrary single logical qubit in another \cubecode\ code block.  Figure~\ref{fig:triple_CNOT_dif_blocks} depicts how to implement the logical $\Lambda(X)^{\otimes 3}$ gate between two \cubecode\ code blocks.  Figure~\ref{fig:CNOT_dif_blocks_circ} depicts an example of how to implement a single logical $\Lambda(X)$ gate between the first logical qubit in one \cubecode\ code block and the second logical qubit in another, utilizing a more elaborate circuit involving intra-block $\Lambda(X)$ gates.

\subsubsection{Transversal Code Switching}

\begin{figure}[!ht]
    \centering
    \begin{align*}
      {\includegraphics[width=0.21\textwidth]{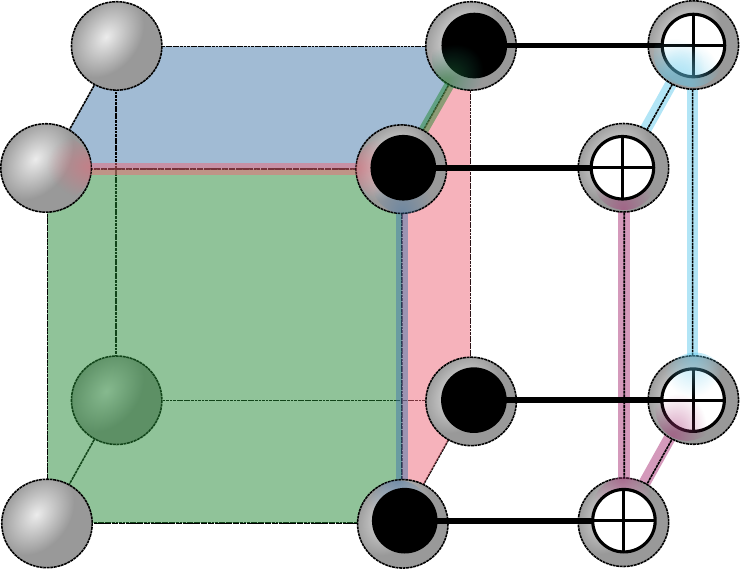}}
      \qquad \raisebox{8ex}{$=$} \qquad 
      \raisebox{8ex}{$\Lambda(X)^{\otimes 2}\left(\rule{0em}{10ex}\right.$
      \raisebox{-8ex}{\includegraphics[width=0.21\textwidth]{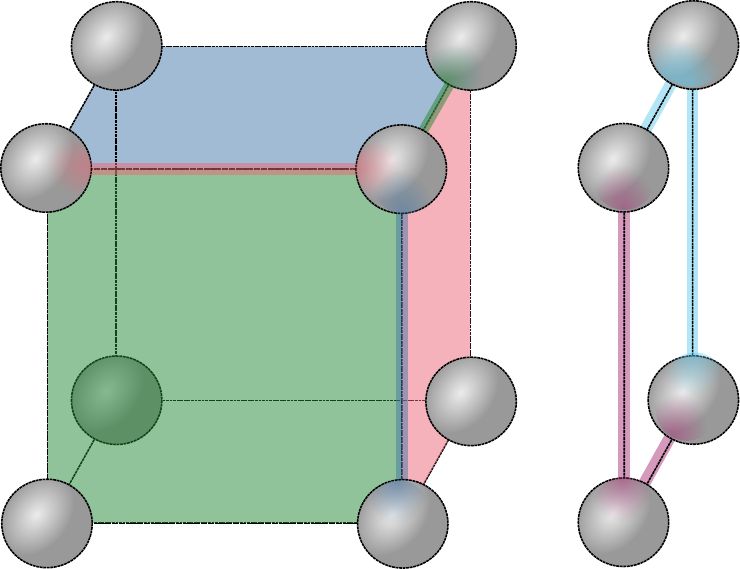}}
      $\left.\rule{0em}{10ex}\right)$}
    \end{align*}
    \caption{Logical $\Lambda(X)^{\otimes 2}$ gate from two control logical qubits in the \cubecode\ code block to the two target logical qubits in the \squarecode\ code block. This is enacted by applying physical $\Lambda(X)$ gates from each physical qubit on a face of the \cubecode\ code block to each physical qubit in the \squarecode\ code block. The face of the \cubecode\ code block is chosen so that its support fully contains both of the participating logical ${Z}$ operators.  Because every possible pair of logical qubits in an \cubecode\ code block is represented on one of its faces, the logical $\Lambda(X)^{\otimes 2}$ gate can be enacted for any desired pair of control logical qubits. This figure depicts logical qubits 3 (blue) and 1 (green) on the \cubecode\ code block as the two control logical qubits.  }
    \label{fig:cube_square_cnot}
\end{figure}

To make use of the combined gate set of the two codes, we teleport states between them. 
To fault-tolerantly apply a transversal logical $\Lambda(X)^{\otimes 2}$ gate from two logical control qubits in the \cubecode\ code block to the two logical target qubits in the \squarecode\ code block, we apply physical $\Lambda(X)$ gates from the four physical qubits in the \cubecode\ code block on the face of the \cubecode\ code block containing the full support of the two relevant logical $Z$ gates to the four physical qubits in the \squarecode\ code~\cite{Wang_2024}.
An example of this in shown in Figure~\ref{fig:cube_square_cnot}.

\begin{figure}[!ht]
    \centering
    \includegraphics{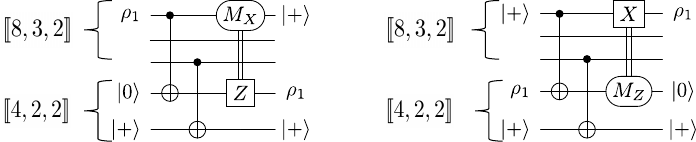}

\caption{Logical teleportation circuits showing $X$-type (left) and $Z$-type (right) teleportation of a single state using the $\Lambda(X)^{\otimes2}$ gate between \cubecode\ and \squarecode\ code blocks. $M_{X/Z}$ denotes a projective measurement onto the $+1$ eigenstate of the measured operator. Both circuits leave the known input qubit state in the initial state for the other type of teleportation, up to a Pauli correction that can be adaptively applied transversally or tracked classically. These circuits can be straightforwardly adapted to teleport two states simultaneously. \label{fig:teleport_circuits}}
\end{figure}

\begin{figure}[!ht]
    \begin{align*}
      \raisebox{1ex}{\includegraphics[width=0.2\textwidth]{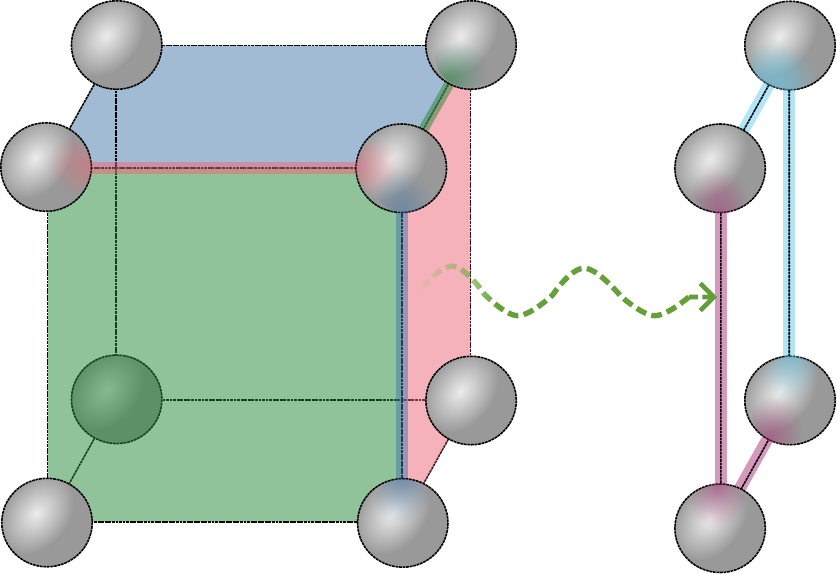}}
      \qquad \raisebox{8ex}{=} \qquad 
      \raisebox{-4ex}
      {\includegraphics[width=0.4\textwidth]{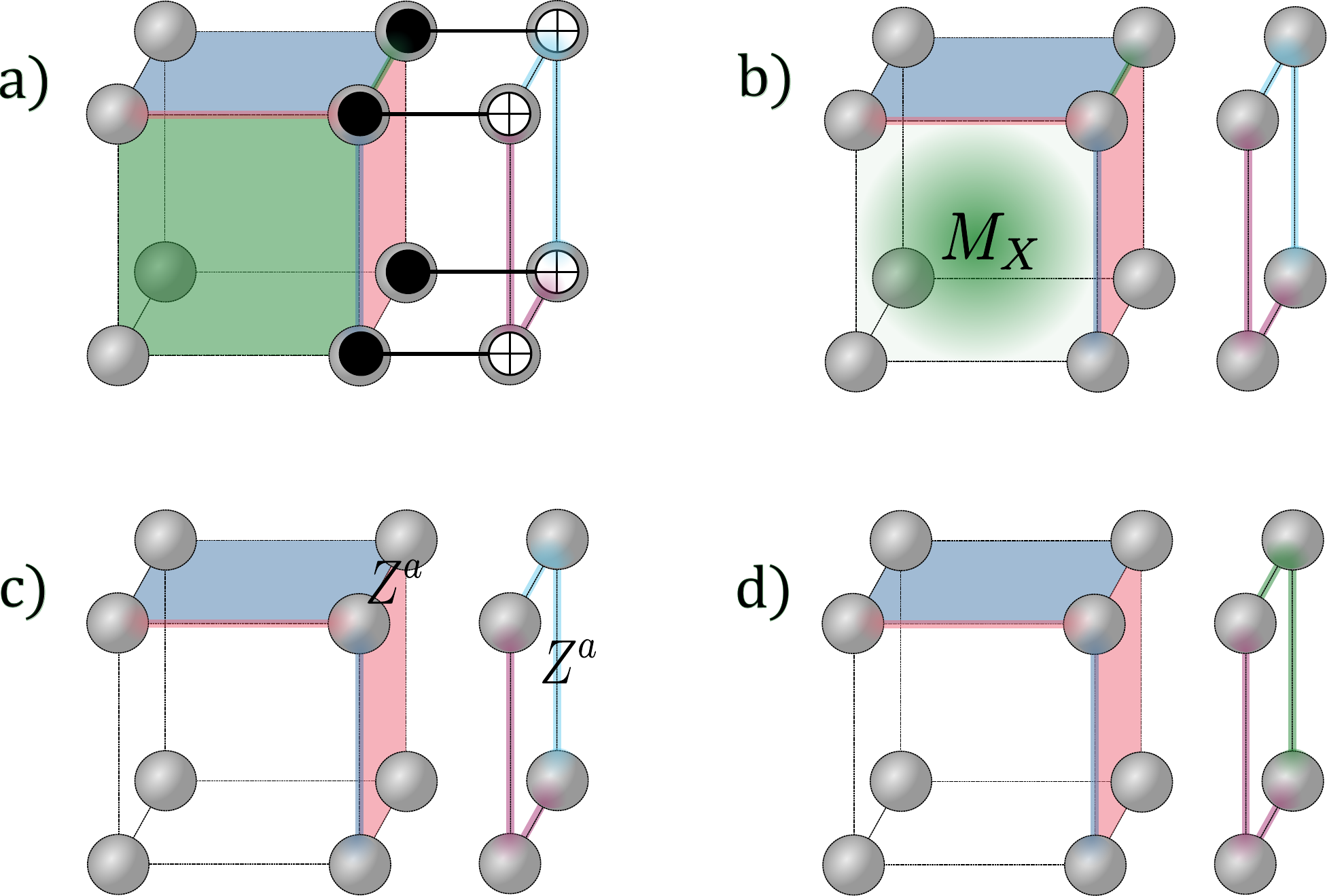}}
    \end{align*}
    \caption{(Left) Tracking the $X$-type teleportation of a single logical qubit  in an \cubecode\ code block (logical qubit 1, green, from Table~\ref{tab:codes}) to a single logical qubit in a \squarecode\ code block (logical qubit 1, light blue, from Table~\ref{tab:codes}).  The two logical qubits in the \squarecode\ code block are initially prepared in the state $|\overline{0}\>_1|\overline{+}\>_2$, while the first logical qubit in the \cubecode\ code is initially in the state $\overline{\rho}_1$. (Right) Broken into steps, one first  a) applies a transversal logical $\Lambda(X)_{11} \otimes \Lambda(X)_{32}$ from the \cubecode\ code block to the \squarecode\ code block.  Then one b) uses an ancilla-coupled measurement protocol, as described in Appendix~\ref{sec:ft_measurements}, to measure the logical ${X_1}$ operator in the \cubecode\ code block, obtaining the outcome $(-1)^a$.  After that, one c) applies physical $Z^a$ gates on the two qubits in the \squarecode\ code block that are in the support of its logical $Z_1$ operator, which effectively implements the logical $Z_1$ gate on the code block fault tolerantly; alternatively these gates could be tracked classically.  At the end, one has  d) teleported logical qubit $1$ in the \cubecode\ code block to logical qubit $1$ in the \squarecode\ code block.}
    \label{fig:X-teleport}
\end{figure}

\begin{figure}[!ht]
    \begin{align*}
      \raisebox{1ex}{\includegraphics[width=0.2\textwidth]{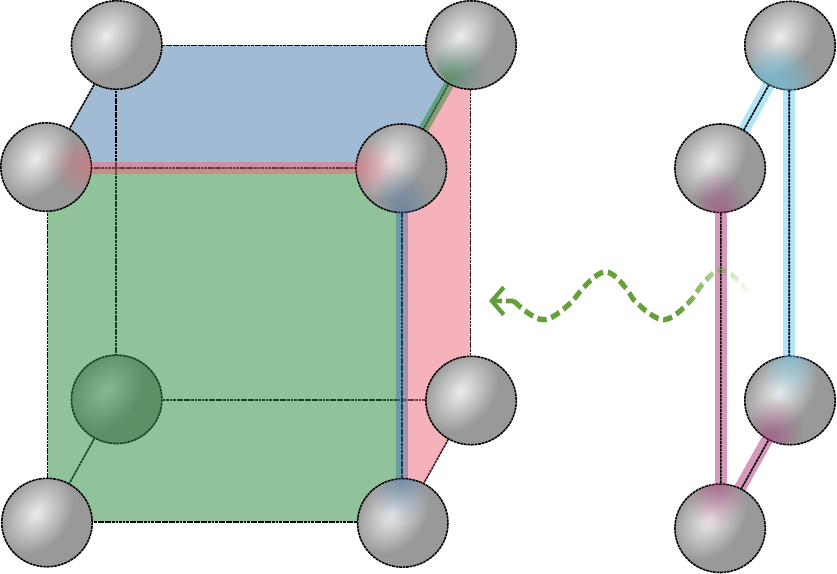}}
      \qquad \raisebox{8ex}{=} \qquad 
      \raisebox{-4ex}
      {\includegraphics[width=0.4\textwidth]{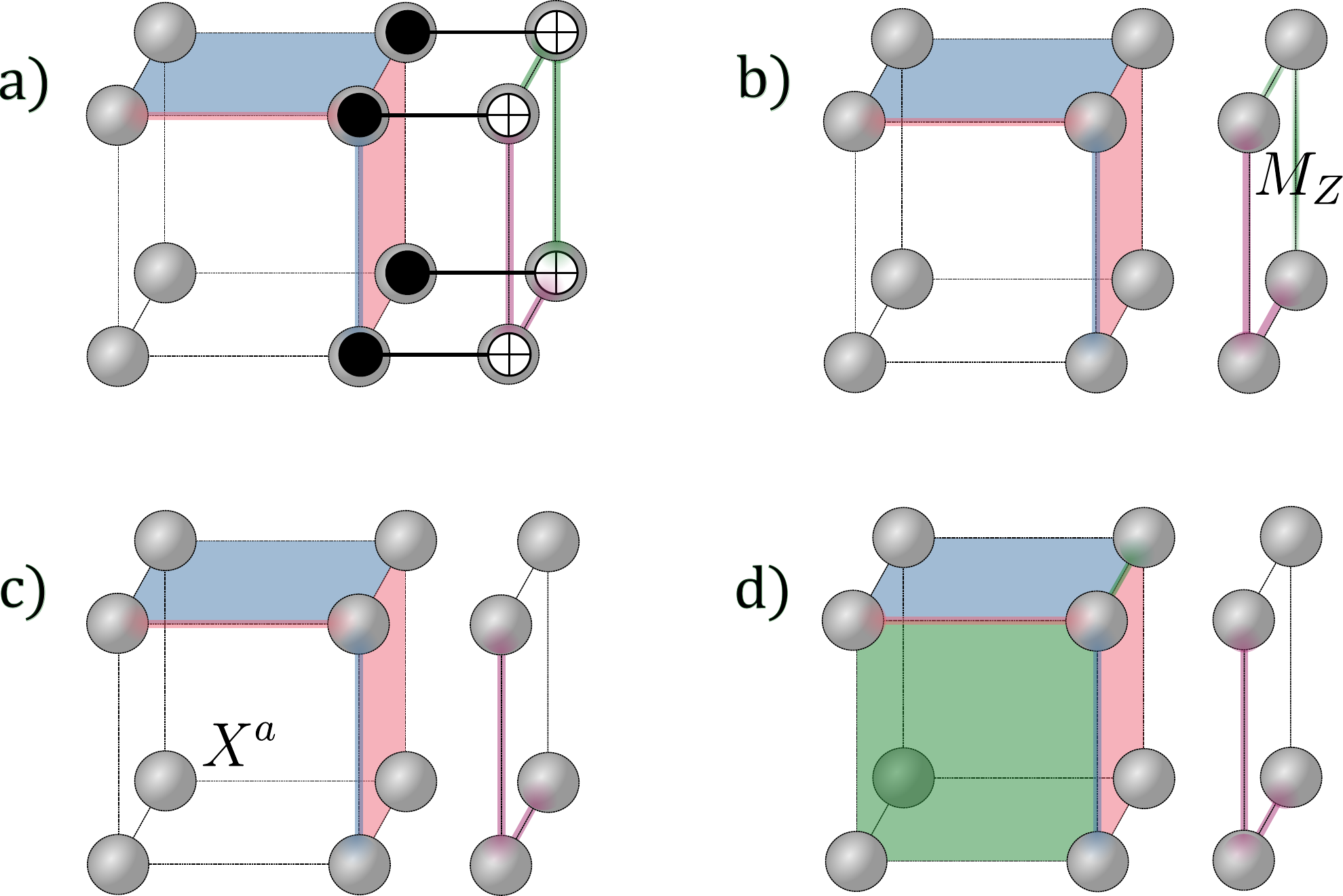}}
    \end{align*}
    \caption{(Left) Tracking the $Z$-type teleportation of a single logical qubit in a \squarecode\ block (logical qubit 1, light blue, from Table~\ref{tab:codes}) to a single logical qubit in an \cubecode\ block (logical qubit 1, green, from Table~\ref{tab:codes}).  The two logical qubits in the \squarecode\ code block are initially in the state $\overline{\rho}_1 \otimes |\overline{+}\>\<\overline{+}|_2$, while the first logical qubit in the \cubecode\ code block is initially prepared in the state $|\overline{+}\>$, which could have happened by as a result of a previous $X$-type teleportation of the logical qubit from the \cubecode\ to the \squarecode.  (Right) Broken into steps, one first  a) applies a transversal logical $\Lambda(X)_{11}$ from the \cubecode\ code block to the \squarecode\ code block.  Then one  b) measures the logical $\overline{Z_1}$ operator in the \squarecode\ code block, obtaining the outcome $(-1)^a$, using an ancilla-coupled measurement protocol described in Appendix~\ref{sec:ft_measurements}.  After that, one c) applies $X^a$ gates on the four qubits in the \cubecode\ code block that are in the support of its logical $X_1$ operator, which effectively implements the logical $X_1$ gate on the code block fault tolerantly, alternatively these gates could be tracked classically.  At the end, one has d) teleported logical qubit $1$ in the \squarecode\ code block to logical qubit 1 in the \cubecode\ code block.}
    \label{fig:Z-teleport}
\end{figure}

As shown in Figure~\ref{fig:teleport_circuits}, and expanded in Figures~\ref{fig:X-teleport} and~\ref{fig:Z-teleport}, a transversal logical $\Lambda(X)$ gate from an \cubecode\ code block to a \squarecode\ code block, followed by transversal logical $M_X$ (resp.~$M_Z$) measurements and logical $|0\>$ (resp.~$|+\>$) preparations, facilitates a logical $X$-type (resp.~$Z$-type) one-bit teleportation circuit~\cite{Zhou:2000a} between the code blocks in a transversal way.\footnote{The fault-tolerant method for transversally measuring only a subset of logical qubits in an $\cubecode$\ or $\squarecode$\ code block is subtle and involves introducing ancillary qubits; it is described in detail in Appendix~\ref{sec:ft_measurements}.}
Because the transversal $\Lambda(X)^{\otimes 2}$ gate between code blocks requires the controls to be on the \cubecode\ block, we use $X$-type teleportation to teleport states from an \cubecode\ code block onto a \squarecode\ block and $Z$-type teleportation to teleport states in the other direction.
To teleport a single-qubit state from an \cubecode\ code block to a \squarecode\ code block or vice-versa, the other qubit on the \squarecode\ code not participating in the teleportation should be initialized to the logical $\ket{{+}}$ state so that the $\Lambda(X)$ gate has no effect.
Likewise, to teleport two states simultaneously, both qubits in the \squarecode\ code should be initialized to the logical $\ket{0}$ state.

For both the \cubecode\ and the \squarecode\ codes, stabilizer-generator measurements and logical-operator measurements can be performed simultaneously to enact the fault-tolerant non-destructive measurements used for the teleportations.
This can be done, for example, by measuring the logical operator followed by measuring the logical operator multiplied by a stabilizer element of the same Pauli type (\eg, one face of the \cubecode\ code and then the opposite face).
If the two measurements yield the same result, the stabilizer element is inferred to be $+1$.
Examples of circuits for these measurements are shown in Ref.~\cite{Wang_2024} and reproduced in Appendix~\ref{sec:ft_measurements}.
This technique can also used to prepare the logical $|0\>$ and $|+\>$ states, by projectively measuring the corresponding logical Pauli operator.

\begin{figure}[!ht]
    \centerline{
    \includegraphics[width=0.5\textwidth]{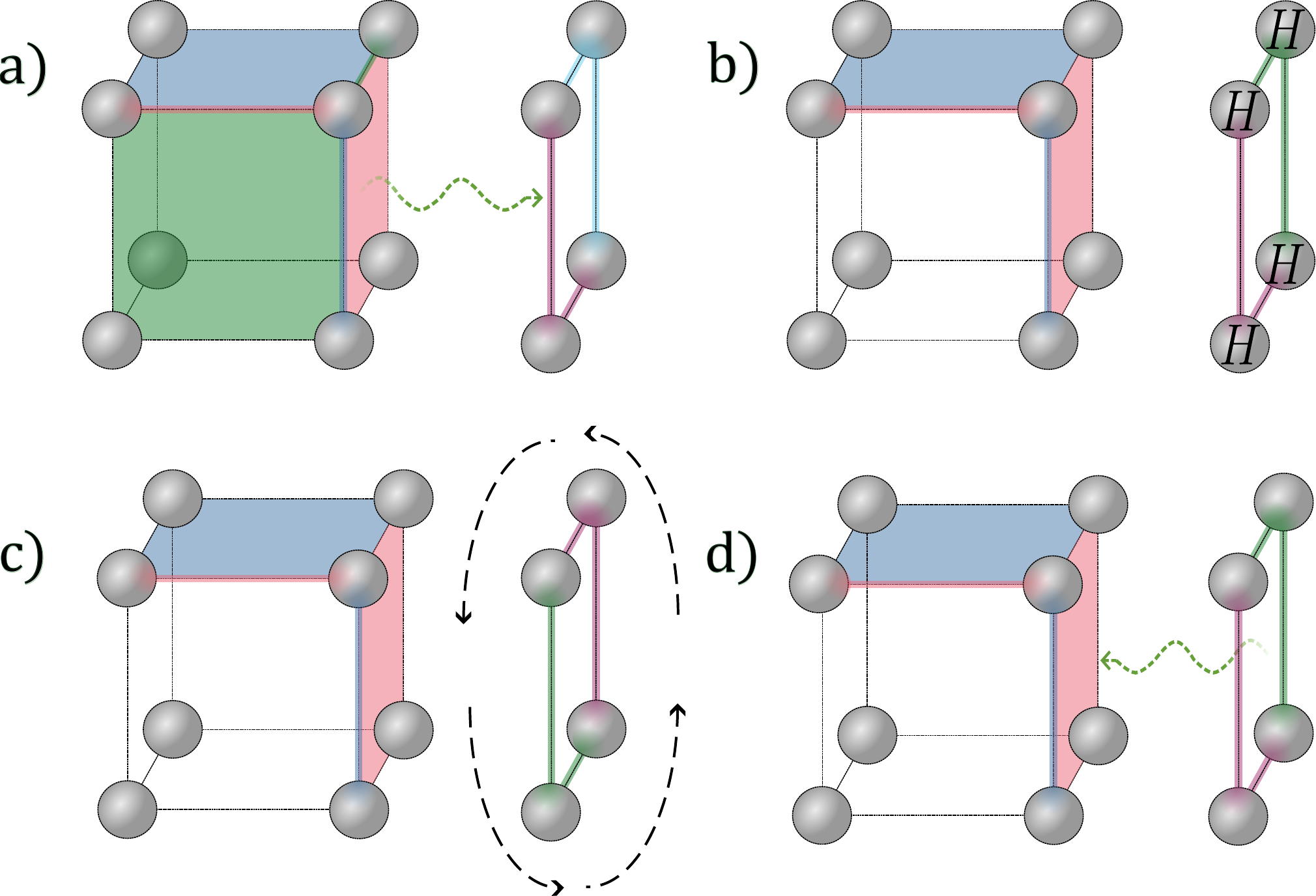}
    }
    \caption{Steps involved in the teleportation-based logical Hadamard gate on a single (green) logical qubit in an \cubecode\ code block. (a) The (green) logical qubit to which the Hadamard will be applied is teleported into a (light blue) logical qubit in a \squarecode\ code block, using $X$-type teleportation, as described in Figure~\ref{fig:X-teleport}.  (b) Transversal Hadamard gates are applied to both qubits in the \squarecode\ code block, while simultaneously swapping the teleported (green) qubit with the the other (red) qubit in the \squarecode\ code block. (c) The $\SWAP$ gate is undone via physical qubit permutations/relabeling. (d) The (green) qubit is teleported back to the \cubecode\ code block. This protocol can be used to enact either one or two logical Hadamard gates at a time, by teleporting one or two logical qubits at a time.}
    \label{fig:teleport_H}
\end{figure}

This teleportation primitive allows for the full universal gate-set to be accessed transversally.
For example, to implement a logical Hadamard gate on a qubit encoded in the \cubecode\ code block, which does not support a direct transversal logical Hadamard gate implementation, we teleport the qubit from the the \cubecode\ code block onto the \squarecode\ code block using $X$-type teleportation. 
Once the state is on the \squarecode\ code block, we apply a transversal Hadamard gate.
We then teleport the qubit back onto the \cubecode\ code block from whence it came using $Z$-type teleportation.
This sequence is shown in Figure~\ref{fig:teleport_H}.
As noted in Sec.~\ref{sec:transversal-gates}, the transversal physical Hadamard gate on the \squarecode\ code block actually enacts a logical Hadamard gate on each of its encoded qubits, followed by logical $\SWAP$ gate.
This can be used to perform two logical Hadamard gates in parallel, by teleporting two states at a time from the \cubecode\ code block instead of just one.  
It also means that in order to teleport the correct qubit back, one must account for the logical $\SWAP$ gate.  
To avoid interfering with other logical qubits in the \cubecode\ code block, the logical $\SWAP$ gate is performed in the \squarecode\ code block before the qubit is teleported back.

\subsubsection{Costs}
\label{sec:costs}

We end this section with a discussion of the relative costs of each operation listed above.
All logical gates executed inside a single block require only physical single-qubit gates or passive permutations, and are thus relatively cheap.
$\Lambda(X)$ gates between logical qubits in distinct code blocks are more expensive, requiring eight physical two-qubit gates to perform logical $\Lambda(X)^{\otimes 3}$ or $16$ physical two-qubit gates to perform targeted $\Lambda(X)$ gates between fewer than three logical qubits at a time.
Finally, teleportation of states between code blocks involves four physical two-qubit gates and ancilla-assisted measurements (shown in Figures~\ref{fig:ft_X} and~\ref{fig:ft_Z}), which involve either $12$ physical two-qubit gates for $X$-type teleportation or eight physical two-qubit gates for $Z$-type teleportation.
Furthermore, each type of teleportation requires four physical qubit measurements.
This makes teleportation the most expensive operation in our architecture. 
These teleportations would not add to the logical depth, whereas these operations add significantly to the physical depth.
This highlights a discrepancy between logical and physical circuit depths; two circuits of the same logical depths may have drastically different physical depths if one requires significantly more teleportations than the other.

\section{Compilation into universal gate sets}
\label{sec:compiler}

\begin{figure}[!ht]
    \centering
    \begin{align*}
        \raisebox{-3ex}{%
        \Qcircuit @C=1em @R=1em {
             & \gate{\underline{S}} & \qw} 
        } 
        \qquad
        \raisebox{-4ex}{=}
        \qquad
        \raisebox{-1ex}{
        \Qcircuit @C=1em @R=1em {
        \lstick{p} & \ctrl{1} & \targ & \qw \\
        & \control \qw  &  \ctrl{-1} &  \qw \\ }
        } 
\end{align*}
    \caption{Decomposition of $S$ gate into the $\{CCZ, \Lambda(Z), \Lambda(X), Z, X, H\}$ gate set. The phase-reference qubit
    is labeled $p$.}
    \label{fig:S_CZCX}
\end{figure}
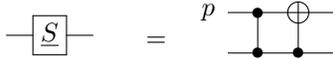

Some quantum algorithms can be mapped to quantum circuits over the $\{CCZ, H\}$ gate set exactly.
However, most utilize transformations that only compile into the $\{CCZ, H\}$ gate set approximately, albeit with an error that can be controlled to exponential precision with polynomial-time effort.
More concretely, the Solovay-Kitaev theorem~\cite{Solovay:1994a, Solovay:2000a, Kitaev:1997b, Kitaev:2002a, Dawson:2005a} asserts that, given a universal gate set $\calB$, a unitary transformation $U$, and a desired precision $\epsilon$, that one can construct a quantum circuit over $\calB$ of size $\textrm{polylog}(1/\epsilon)$ in a time $\textrm{polylog}(1/\epsilon)$ that is within a distance $\epsilon$ of $U$.
Numerous ``quantum compiling'' or ``quantum gate synthesis'' algorithms for constructing approximating quantum circuits in this way have been developed, including a highly optimized one over the $\{\textrm{Clifford}, \Lambda(S)\}$ gate set by Glaudell, Ross, and Taylor~\cite{glaudell2021optimaltwoqubitcircuitsuniversal}.
The authors also provide a Mathematica program that implements their synthesis algorithm that is freely available for download~\cite{andrewglaudell_2021_4549819}.
In the typical case, their algorithm yields circuit approximations whose gate counts are 
\begin{align}
    T(\epsilon) &= 5 \log_2(1/\epsilon) + \bigO(1).
\end{align}
Because $\Lambda(S)$ can be exactly simulated with the $\{CCZ, H\}$ gate set within the phase-reference encoding, this algorithm can be boot-strapped to the $\{CCZ, H\}$ gate set by first approximately synthesizing into the $\{\textrm{Clifford}, \Lambda(S)\}$ gate set and then exactly synthesizing all of those gates into their phase-reference encoded equivalents using only the $\{CCZ, H\}$ gate set.
Recall that the Clifford group is generated by the $\{H, S, \Lambda(X)\}$ gate set.  
Since we have already shown that we have access to logical $\Lambda(Z)$, $\Lambda(X)$, $X$, and $Z$ gates on the \cubecode\ and \squarecode\ codes, we can immediately expand our exactly synthesizable logical gate set to $\{CCZ, \Lambda(Z), \Lambda(X), Z, X, H\}$.
Now the only logical gates to convert to the phase-reference encoding within the generating set for the $\{\textrm{Clifford}, \Lambda(S)\}$ gate set are the logical $\Lambda(S)$ and $S$ gates.
We have already shown how to exactly synthesize the $\Lambda(S)$ gate using this gate set in Figure~\ref{fig:CS}.  
To exactly syhtesize the $S$ gate, one can remove a single control qubit from this construction, as is depicted in Figure~\ref{fig:S_CZCX}.

The synthesis algorithm of Reference~\cite{glaudell2021optimaltwoqubitcircuitsuniversal} is limited to two-qubit ($SU(4)$) unitary gates.
However, for our purposes, that is sufficient.
Every multi-qubit gate can be decomposed exactly into single-qubit rotations and $\Lambda(X)$ gates, which are contained within $SU(4)$.
The restriction this poses is that for a circuit to be compiled to our architecture, it must first be expressed using only two-qubit and $\CCZ$ gates (or exact-Clifford-synthesis-equivalent $\CCZ$ gates).

One may wonder if a similar strategy is possible by first synthesizing into the common $\{\textrm{Clifford}, T\}$ gate set and then converting into the $\{CCZ, H\}$ gate set.
This is impossible because the $T$ gate cannot be exactly synthesized with the $\{CCZ, H\}$ gate set~\cite{Amy_2020,mukhopadhyay2024synthesizingtoffolioptimalquantumcircuits}.
However, if allowed access to a single magic state of a suitable type, the $T$ gate can be realized with the $\{CCZ, H\}$ gate set in a catalytic way~\cite{amy2023catalyticembeddingsquantumcircuits}, namely in a way in which the magic state is not consumed in the process and can be reused.
In that sense, the $\{\textrm{Clifford}, T\}$ gate set can be catalyzed by $\{CCZ, H\}$ gate set up to the fidelity of the magic state provided~\cite{kissinger2024catalysingcompletenessuniversality}.
We describe one such construction in Appendix~\ref{sec:catalysis}.

\section{Circuit execution}
\label{sec:architecture}

Having shown how to implement a universal gate set in the \cubecode\ and \squarecode\ codes, and how to compile arbitrary circuits into that gate set, we now explain the execution of such circuits in detail. 

\subsection{Step 0: Allocate space}
\label{sec:space}

\begin{figure}[!ht]
    \centering
    \includegraphics[width=0.35\textwidth]{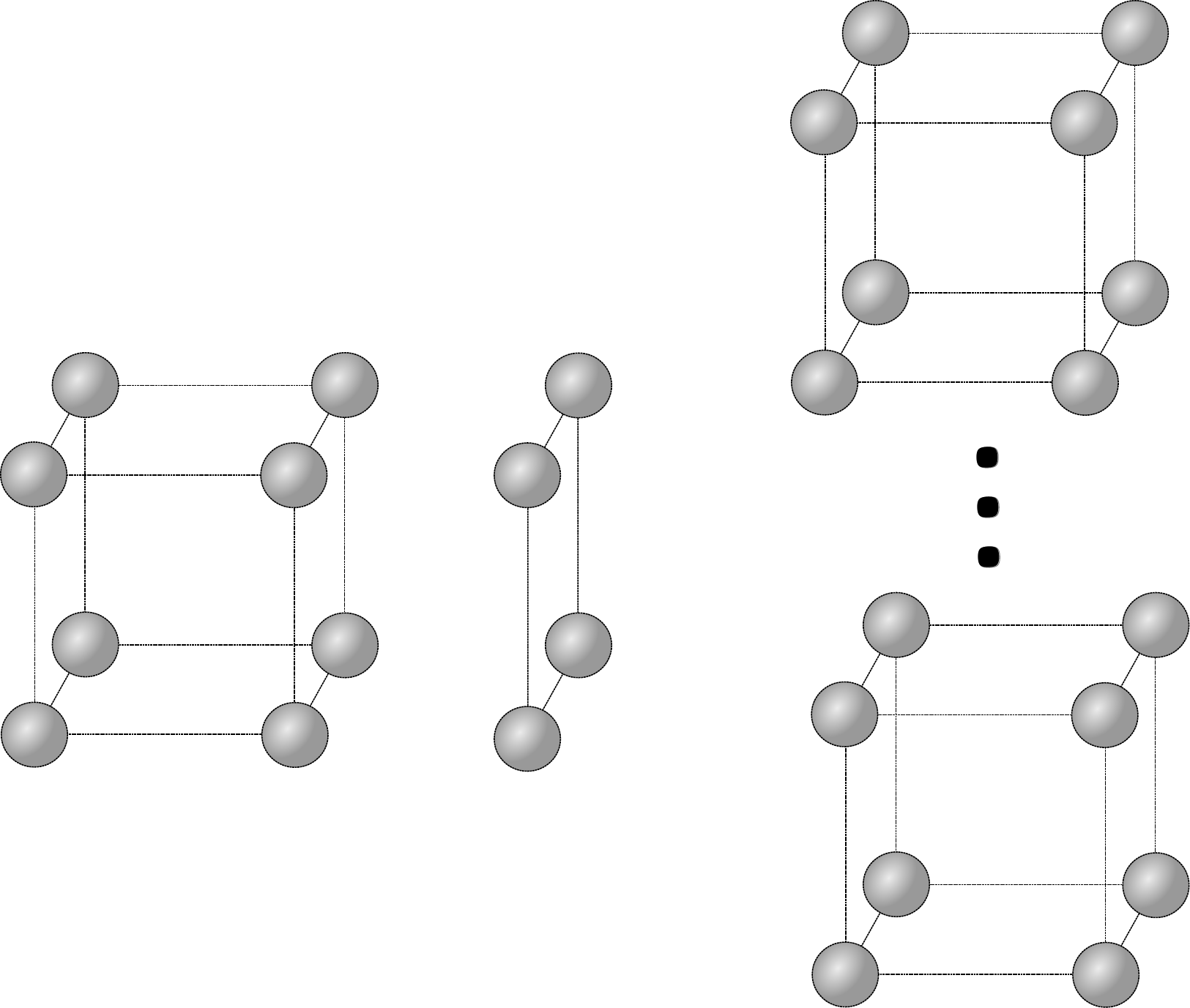}
    \caption{Conceptual layout of the logical architecture. The \cubecode\ blocks on the right are the data blocks, the \cubecode\ block on the left is the rotation block, and the \squarecode\ block is the transfer block.
    This layout is repeated in Figures~\ref{fig:enter-label}--\ref{fig:Log_Rot}. The ancilla qubits required to perform fault-tolerant stabilizer measurements are not displayed here.}
    \label{fig:log_space}
\end{figure}

Before execution begins, there must be sufficient resources to encode the logical qubits identified in the circuit itself (\textit{i.e.}, $n_c$ ``computational qubits''). 
Thus, there must be $\lceil n_c/3 \rceil$ \cubecode\ blocks, which we will refer to as ``data blocks,'' to store the computational qubits.
We will need one additional \cubecode\ block, which we will call the ``rotation block,'' to store the phase-reference qubit.
During the course of the computation, the remaining two qubits of the rotation block will be used to temporarily store the states of computational qubits.
Finally, we need at least one \squarecode\ block to perform Hadamard gates and to transfer states between the rotation and data blocks.
We will refer to these \squarecode\ blocks as ``transfer blocks.''
These are illustrated in Figure~\ref{fig:log_space}.

In addition to the physical space required to realize the data, rotation, and transfer blocks, ancilla qubits are required in order to facilitate fault-tolerant stabilizer measurements.
As described in Appendix~\ref{sec:ft_measurements}, each stabilizer generator of both the \cubecode\ and \squarecode\ codes can be measured fault-tolerantly using two ancilla qubits.
If the physical architecture admits a high degree of reconfigurability, these two ancilla qubits could in principle be used for all stabilizer measurements on all code blocks.
This would come at the cost of requiring that stabilizer measurements be performed sequentially, rather than in parallel.  
Although scalable fault-tolerant designs require parallel syndrome extraction circuits, this early FTQC architecture is fixed at distance two, so this is not a consideration.
These two qubits, along with the qubits needed for the data, rotation, and transfer blocks, leads to a requirement that our architecture have at least the following number of physical qubits available for logical execution on $n_c$ logical qubits:
\begin{equation}
    n_{p,\text{min}} = 8 \Bigl\lceil \frac{n_c}{3} \Bigr\rceil + 14.
\end{equation}

Alternatively, we can increase the number of ancilla qubits in order to perform stabilizer measurements in parallel.
The maximum degree of parallelization would be to simultaneously measure all stabilizer generators at once.
This would require $10$ physical qubits per \cubecode\ block---$2$ for each of its $5$ stabilizer generators---and $4$ physical qubits per \squarecode\ block---$2$ for both of its stabilizer generators.
In total, this requires 

\begin{equation}
    n_{p,\text{fast}} = 18 \Bigl\lceil \frac{n_c}{3} \Bigr\rceil + 26
\end{equation}
physical qubits for logical execution.

A reasonable middle ground would be to add $2$ ancilla qubits per \cubecode\ and \squarecode\ block. 
The stabilizer generators of each individual block would be measured sequentially, however, this would be done in parallel over all distinct blocks.
This would ensure the circuit depth for a full round of stabilizer measurements is a constant that is independent of how many total blocks there are.
This set up would require

\begin{equation}
    n_{p,\text{mid}} = 10 \Bigl\lceil \frac{n_c}{3} \Bigr\rceil + 18
\end{equation}
physical qubits for logical execution.

\subsection{Step 1: Initialize}

\begin{figure}[!ht]
    \centering
    \includegraphics[width=0.35\textwidth]{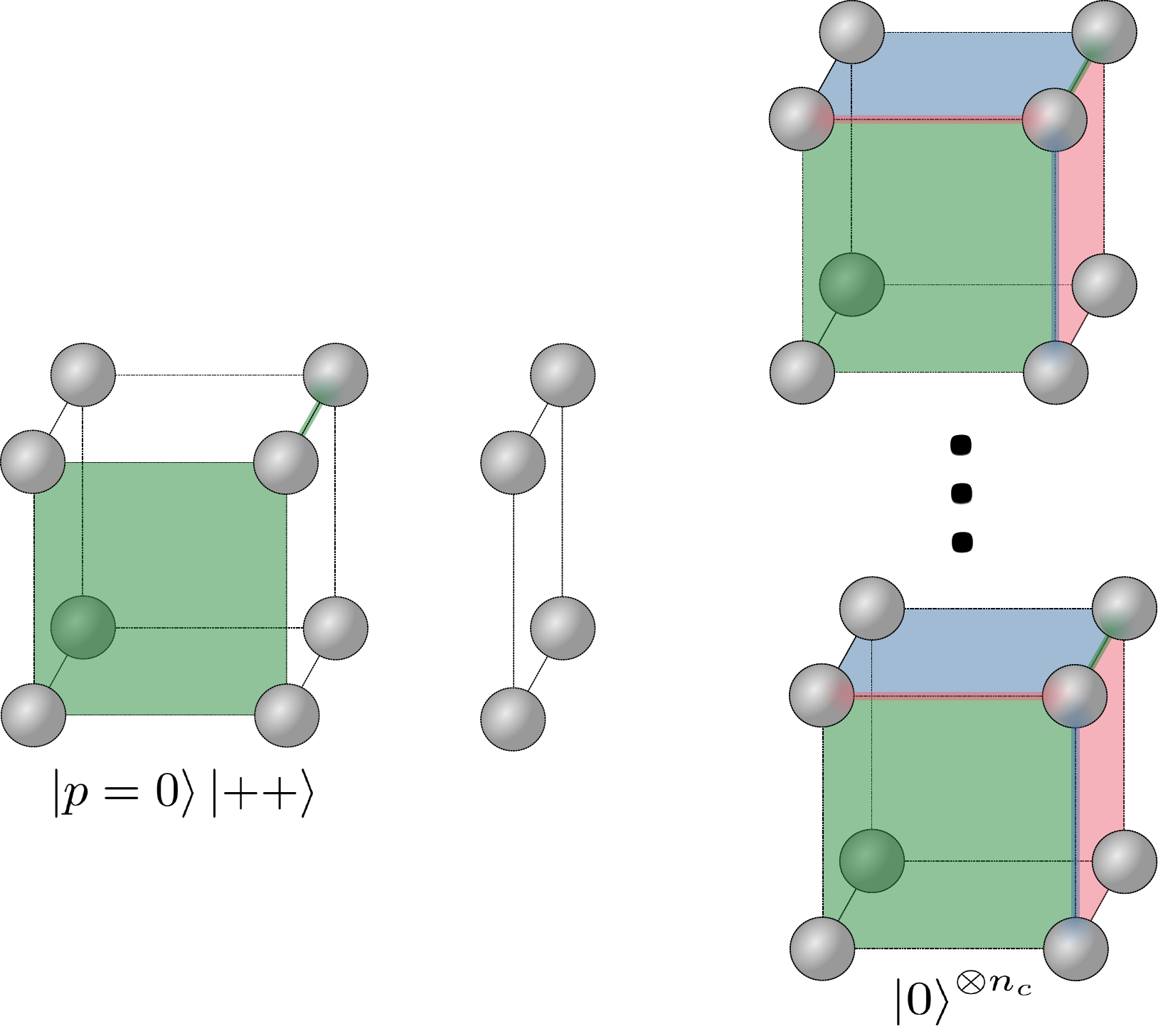}
    \caption{Computation begins by initializing all computational logical qubits and the phase-reference logical qubit to the $\ket{0}$ state. The remaining two logical qubits in the rotation block are initialized to $\ket{+}$ states as targets for teleportation operations.}
    \label{fig:enter-label}
\end{figure}

Execution begins with the initialization of the phase-reference logical qubit and all computational logical qubits to the $\ket{{0}}$ state.
The remaining two logical qubits in the rotation block are initialized into the $\ket{{++}}$ state to serve as targets for future teleportation operations.  
The conceptual layout of this step is depicted in Figure~\ref{fig:enter-label}.

\subsection{Step 2: Execute all gates}

The first step in the compilation method given in Section~\ref{sec:compiler} is to express the computation as a sequence consisting of $SU(4)$ and $\CCZ$ gates.
These $SU(4)$ gates are then synthesized to quantum circuits over the $\{\CCZ, H, \Lambda(X), \Lambda(Z), Z, X\}$ gate set.
These synthesized gates will require the target qubits being in the same \cubecode\ block as the phase-reference.
As such, each synthesized $SU(4)$ gate will be performed by teleporting the specific qubits through the transfer block and into the rotation block.
Once inside the rotation block, all gates for the synthesized $SU(4)$ are executed as specified in Section~\ref{sec:422-832-logical-gates}.
The qubits are then teleported back into their original data blocks.
When the two computational qubits originate from separate data blocks, they must be retrieved one at a time, thus requiring twice as many teleportation operations as when the computational qubits originate from the same data block.
Because of this, it is optimal to lay out the qubits so the number of synthesized $SU(4)$ gates performed on computational qubits on separate data blocks is minimized.
Gates containing only real components that are not part of a synthesized $SU(4)$ gate, such as $\Lambda(X)$ and $\Lambda(Z)$, do not alter the phase-reference and can therefore be implemented directly in the data blocks without requiring teleportation into the rotation block.

\begin{figure}[!ht]
    \centering
    \includegraphics[width=0.8\linewidth]{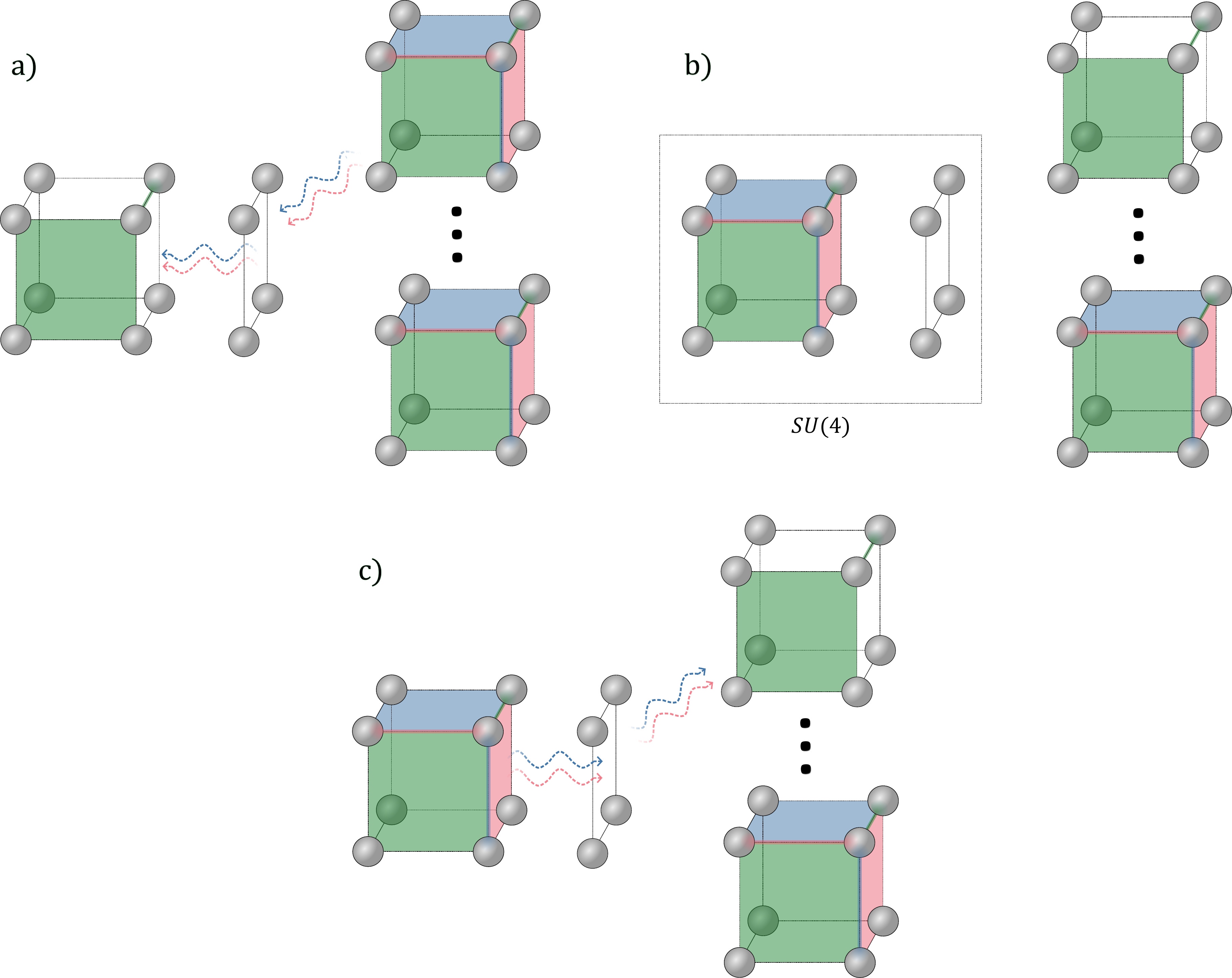}

    \caption{A logical $SU(4)$ gate is performed on a pair of computational qubits by first a) teleporting the states through the transfer block and into the rotation block, where they have access to the phase reference qubit, b) performing all the gates in the synthesized rotation between the rotation block and the transfer block, and c) teleporting the states through the transfer block and back to the data block from whence they originated. The states shown here originate from the same data block. A synthesized $SU(4)$ gate can be performed on qubits from different data blocks by retrieving the states one at time, performing twice as many teleportations as when the states come from the same data block.}
    \label{fig:Log_Rot}
\end{figure}

\subsection{Step 3: Measure logical operators}

The computation concludes after all gates are completed.
At that point, the logical $Z$ operators or $X$ operators, as prescribed by the specific computation, are measured.
As there is no dependence on the phase-reference qubits, the measurements are performed in the data blocks.

If only logical $Z$ operators are measured, all qubits in the \cubecode\ block may be measured destructively in their $Z$ bases and the logical $Z$ operators and $Z$ stabilizer generators can be reconstructed from the measurement outcomes.
Likewise, if only logical $X$ operators are measured, all qubits in the \cubecode\ block may be measured destructively in their $X$ bases with the logical $X$ operators and $X$ stabilizer generators reconstructed from the measurement outcomes.
If both logical $Z$ and logical $X$ measurements are needed in a single block, the measurements must be non-destructive.
This can be accomplished using the measurement scheme utilized for the teleportation operations, as described in Appendix~\ref{sec:ft_measurements}.

\section{Performance Characterization}
\label{sec:performance}

We have now shown how to compile and execute arbitrary computations in our architecture.
However, because it uses QEC codes at distance two, we expect it to be limited to relatively short circuits before noise overwhelms any benefit that error detection and post-selection might confer.
Thus, we would like to be able to characterize the types of circuits that a device realizing this architecture can implement with high fidelity, and to understand the extent to which those capabilities advance with improvements in the underlying physical hardware.
In this section, we sketch a protocol to do exactly that.

Volumetric benchmarking protocols characterize the capability of a quantum processor in terms of the width and depth of the circuits that it can execute with a high probability of success.
So far, these protocols have largely focused on notions of width and depth defined in terms of the number of physical qubits and layers of gates applied to them, respectively.
Care must be taken in defining these in a logical architecture, particularly for EFTQCs.
For example, in Section~\ref{sec:space} we showed that different choices for stabilizer measurement led to different physical qubit requirements for the same number of logical qubits.
As such, specifying only the number of logical qubits does not fully specify the number of physical qubits.
Likewise, as shown in Section~\ref{sec:costs}, logical teleportations do not add to the logical depth but contribute significantly to the physical depth.
Of course, a logical circuit could be characterized in terms of a physical width and depth, however, doing so would shift the focus from the logical level down to the physical. 
To characterize circuits at the logical level we propose (a) fixing the stabilizer measurement strategy, then (b) reporting the logical width to be the number of logical qubits, then (c) tracking the number of transversal logical gates and logical teleportations individually, and finally (d) assigning a ``hardness'' to the circuit that is equal to the number of teleportations required to complete it.
We then characterize a logical circuit by the stabilizer measurement strategy, logical width, depth, and hardness.

While the QEC codes used in this architecture can detect any single physical error, if more than one physical error occurs it still might be the case that a shot and its attendant measurements are accepted.
Thus for any given shot there are three possibilities: no errors occur, an error occurs and is detected, or an an error occurs and is not detected.
Denoting the number of occurrences of each respective event as $N_0$, $N_{det}$, and $N_{und}$, we define the probabilities:

\begin{subequations}
\begin{align}
    p_0 &= \frac{N_0}{N_0 + N_{det} + N_{und}}, \\
    p_{det} &= \frac{N_{det}}{N_0 + N_{det} + N_{und}},~\text{and} \\
    p_{und} &= \frac{N_{und}}{N_0 + N_{det} + N_{und}}. 
\end{align}
\end{subequations}

Because the logical qubits in this architecture are fixed at distance two, logical error rates are improved through error-detection and post-selection rather than active error correction.
This implies discarding the events where an error is detected.
The fraction of shots which are accepted is then
\begin{equation}
    \frac{N_0 + N_{und}}{N_0 + N_{det} + N_{und}} = 1-p_{det}.
\end{equation}
The probability that an accepted shot is error free is
\begin{equation}
    \frac{N_0}{N_0 + N_{und}} = \frac{p_0}{1-p_{det}}.
\end{equation}
We then define two metrics to assess the performance of a physical device utilizing this architecture to implement a logical circuit: the yield ($\Gamma$) as the probability that a shot survives error-detection and post-selection and the confidence ($C$) as the probability that a post-selected shot is error free,
\begin{align}
    \Gamma &= 1-p_{det}~\text{and} \\
    C &= \frac{p_0}{1-p_{det}}.
\end{align}

Because $C$ is the probability that there were no errors after post-selection, $C$ can be interpreted as a post-selected entanglement fidelity~\cite{schumacher1996sendingquantumentanglementnoisy,nielsen1996entanglementfidelityquantumerror}.
Entanglement fidelity is a measure of how well a quantum channel preserves entanglement and can be seen to be equal to the probability that no errors occur~\cite{proctor2022establishing}.
Mirror circuit fidelity estimation (MCFE)~\cite{Proctor_2022,proctor2022measuring,mayer2023theory} is a benchmarking technique that estimates entanglement fidelities by running experiments on ensembles of circuits whose noiseless implementations resolve to the identity.
We adapt this technique to estimating $C$ and further details of the specifics of mirror circuits can be found in Appendix~\ref{sec:C_est}.

The ensemble of circuits studied in an MCFE experiment is generated by randomized Clifford conjugation of mirror circuits---\emph{i.e.}, implementing single-qubit Clifford twirling.
For our architecture, we will absorb the Clifford layers into state preparation and measurement, which would otherwise be in the computational basis.
Thus, each circuit begins with the preparation of an eigenstate of a $k$-fold tensor product of logical Pauli operators.
This state can be represented as a length-$k$ list specifying which Pauli ($X$, $Y$, or $Z$) each logical qubit is in an eigenstate of and a $k$-bit string ($\Vec{b}$) specifying whether the qubit is in the $+1$ ($0$) or $-1$ ($1$) eigenstate.
After the mirror circuit is performed, each logical qubit is measured in the Pauli basis in which it was initialized. 
In the ideal case, this measurement should reproduce $\Vec{b}$, however, errors may change the outputs to broader classical probability distributions over bit strings.
The deviation from the bit string produced by the noiseless mirror circuit can be used to estimate the entanglement fidelity~\cite{Proctor_2022,proctor2022establishing,UniversalMRB}.
We note that, in contrast to physical MCFE techniques, there is no randomized compiling step required here.
This is because measuring stabilizer generators projects any coherent errors in the circuit to stochastic Pauli errors, so there is less concern that coherent errors will cancel out in the mirror circuit\footnote{It will be the case that these measurements project coherent errors onto stochastic Pauli errors if the measurement circuits behave according to standard assumptions in QEC, but one reason to use a protocol like this is to assess whether those assumptions are met in practice.}.

To be precise, let $M$ be a mirror circuit and $L$ be a circuit layer consisting of single-qubit Clifford gates chosen uniformly at random.
We are concerned with estimating the probability of successfully executing $L^{\dagger}ML$.
Ref.~\cite{proctor2022establishing} defines a success probability that accounts for errors that do not affect the measurement outcome (\emph{e.g.}, a $Z$ error does not cause a bit flip when measuring in the $Z$ basis), 
\begin{equation}
    S(M) = \sum_{j=0}^k \left( -\frac{1}{2} \right)^j h_j(M),
    \label{eq:S(M)}
\end{equation}
where $h_j(M)$ is the probability of observing a bit string with Hamming distance $j$ from the ideal output of $L^{\dagger}ML$.
The expectation value of $S(M)$ over many different instances of $L$ is then equal to the entanglement fidelity ($\mathcal{F}_e$) of $M$,
\begin{equation}
    \langle S(M) \rangle = \mathcal{F}_e(M).
\end{equation}

Having outlined the procedure for MCFE, we can now consider implementing it in our particular architecture.
It is straightforward to absorb the single-qubit Clifford into state preparation and measurement.
Preparation and measurement in the $Z$ and $X$ bases are already primitives of this architecture, but
preparation and measurement in the $Y$ basis is slightly more complicated.
Using $SXS^{\dag} = Y$, the basis states of $Y$ can be prepared by applying the transpiled $S$ gate~\ref{fig:S_CZCX} to basis states of $X$.
Likewise, measurement in the $Y$ basis is performed by applying the transpiled $S$ before measuring in the $X$ basis.
The mirror circuit $M$ can then be implemented according to the steps outlined in Section~\ref{sec:architecture}.
The protocol for estimating $C$ and $\Gamma$ is summarized in Algorithm~\ref{alg:cap} and expanded upon in Appendix~\ref{sec:C_est}.
This protocol could be repeated for many different logical widths, depths, and values of hardness.

\begin{algorithm}[!ht]
  \caption{Protocol for estimating $C$ and $\Gamma$.}\label{alg:cap}
  \begin{algorithmic}[1]
      \State \textbf{input} $\{M_i(w,d,\eta) \}$, a set of $N_m$ mirror circuits of logical width $w$, depth $d$, and hardness $\eta$.\\
      \hspace{3.0em} $N_s$, the number of shots per mirror circuit.
      \State \texttt{confidence} $= 0$
      \State \texttt{yield} $= 0$
      \For{ $M$ in $\{M_i(w,d,\eta) \}$}
        \For{$n$ < $N_s$}
            \State \texttt{hamming\_distances} = \texttt{Zeros}($w$)
            \State \texttt{runs\_kept} $= 0$
            \State $L_n = $ \texttt{RandomCliffordLayer}($w$),
            \State $\ket{\overline{L}_n} = L_n\ket{\overline{0}}$
            \State $\ket{M_L} = M\ket{\overline{L}_n}$
            \If{(\texttt{no\_errors\_detected})}
                \State $b_{M,n} = \texttt{Measure}(\ket{M_L}, \texttt{basis} = L_nZ^{\otimes w}L_n^{\dagger} )$
                \State $q = $ \texttt{HammingDistance}($b_{M,n},\Vec{b}(M)$)
                \State \texttt{Increment}($\texttt{hamming\_distances}[q]$)
                \State \texttt{Increment}(\texttt{runs\_kept})
            \ElsIf{(\texttt{error\_detected})}
                \State \texttt{DiscardRun()}
            \EndIf
        \EndFor
        \texttt{hamming\_distances} = $\frac{\texttt{hamming\_distances}}{N_s}$ 
        \State $\texttt{confidence} \pluseq  \sum_{j=0}^w \left( -\frac{1}{2} \right)^j \texttt{hamming\_distances}[j]$ 
        \State $\texttt{yield} \pluseq \frac{\texttt{runs\_kept}}{N_s}$ 
      \EndFor 
    \State \Return $\frac{\texttt{confidence}}{N_M}$, $\frac{\texttt{yield}}{N_M}$
  \end{algorithmic}
\end{algorithm}

Here, we do not specify which mirror circuits to run.
It would be natural to first characterize the architectural primitives listed in Section~\ref{sec:422-832-logical-gates} (\emph{e.g.}, the logical operations and the teleportation operations) and then  characterize increasingly larger circuits comprised of these primitives.
We expect the quality of the estimates provided by this protocol will be heavily dependent on the selection of the mirror circuits explored.
We leave the determination of optimal mirror circuits for this and other logical architectures for future work. 

\section{Summary and Future Work}
\label{sec:summary_future_work}

We have presented a universal architecture designed for early fault-tolerant quantum computers based on the \cubecode\ and \squarecode\ codes.
We have also introduced a exact transpiliation strategy for converting from the $\{\textrm{Clifford}, \Lambda(S)\}$ gate set to the $\{\CCZ, H\}$ gate set, which was specialized to gates accessible to the \cubecode\ and \squarecode\ codes considered jointly.
Our architecture avoids the need for magic states, instead achieving universality with only the transversal logical gates available on the \cubecode\ and \squarecode\ codes and teleportation operations between them.
Finally, we have proposed a simple logical benchmarking protocol, based on mirror circuit fidelity estimation, to assess the capabilities of a quantum computer using this architecture.

Further work could include generalizing this architecture to codes with distance greater than two, optimizing the selection of sets of mirror circuits for confidence estimation, developing more sophisticated logical benchmarking protocols of fault-tolerant architectures, and implementing this architecture on a hardware platform.

\begin{acknowledgements}
We thank 
Earl Campbell,
Riley Chien,
Cole Maurer, 
Setso Metodi,
Ben Morrison,
Tim Proctor,
Mason Rhodes,
Kenny Rudinger,
Stefan Seritan,
and Jalan Ziyad for helpful comments and technical discussions.
We are also grateful to Setso Metodi for prepublication review and broader administrative and moral support.
We happily acknowledge Paul Tol's online notes on color schemes, \url{https://sronpersonalpages.nl/~pault/}, for inspiring the color palette choices for our figures.

This work is supported by a collaboration between the US DOE and other Agencies. This material is based upon work supported by the U.S.\ Department of Energy, Office of Science, National Quantum Information Science Research Centers, Quantum Systems Accelerator. Additional support is acknowledged from the U.S.\ Department of Energy, Office of Fusion Energy Sciences, Foundations for Quantum Simulation of Warm Dense Matter project.  Specifically, JSN and ADB were supported by DOE FES and AJL was supported by DOE NQIS QSA.

This article was co-authored by employees of National Technology \&
Engineering Solutions of Sandia, LLC under Contract No.\ DE-NA0003525 with the
U.S. Department of Energy (DOE). The authors own all rights, title, and interest
in and to the article and are solely responsible for its contents. The United
States Government retains and the publisher, by accepting the article for
publication, acknowledges that the United States Government retains
a non-exclusive, paid-up, irrevocable, world-wide license to publish or
reproduce the published form of this article or allow others to do so, for
United States Government purposes. The DOE will provide public access to these
results of federally sponsored research in accordance with the DOE Public Access
Plan \url{https://www.energy.gov/downloads/doe-public-access-plan}.

Sandia National Laboratories is a multimission laboratory managed and operated by National Technology and Engineering Solutions of Sandia, LLC., a wholly owned subsidiary of Honeywell International, Inc., for the U.S. Department of Energy's National Nuclear Security Administration under contract DE-NA-0003525.

\end{acknowledgements}

\bibliographystyle{quantum}
\bibliography{references,landahl}

\appendix

\section{Fault-tolerant measurements for logical \texorpdfstring{$X$}{X}-type and \texorpdfstring{$Z$}{Z}-type teleportation}
\label{sec:ft_measurements}

To perform a logical one-bit teleportation operation, a logical non-destructive Pauli $X$ or $Z$ measurement is required, as depicted in Figure~\ref{fig:teleport_circuits}.
The standard way to do this is via an ancilla-coupled scheme in which, for an $X$ (resp.~$Z$) measurement, the ancilla is prepared in the $|+\>$) (resp.~$|0\>$ state) and a sequence of $\CNOT$ gates whose targets (resp.~controls) comprise the support of the logical operators in the code block and whose control (resp.~target) is on the ancilla qubit, after which the the ancilla qubit is measured in the $X$ (resp.~$Z$) basis.

Because errors could occur during this process, it must be wrapped in some kind of fault-tolerant design to maintain the level of error suppression afforded by the codes involved.  
A solution to this is described in Ref.~\cite{Wang_2024}, inspired by flag-fault-tolerant syndrome extraction, as described in Ref.~\cite{Chao:2020a}.  
In the solution, essentially two forms of redundancy are utilized.  

\begin{figure}[!ht]
    \centering
    \includegraphics{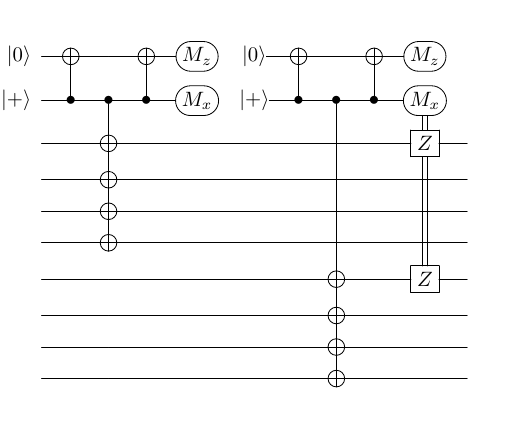}
    \caption{Fault-Tolerant $X_1$ measurement in the \cubecode\ code with conditional $Z$ corrections. No error is detected if both $Z$measurements return $0$ and both $X$ measurements return the same result. For the teleportation scheme the conditional $Z$ corrections are also applied to the qubits in the \squarecode\ code.}
    \label{fig:ft_X}
\end{figure}

\begin{figure}[H]
    \centering
    \includegraphics{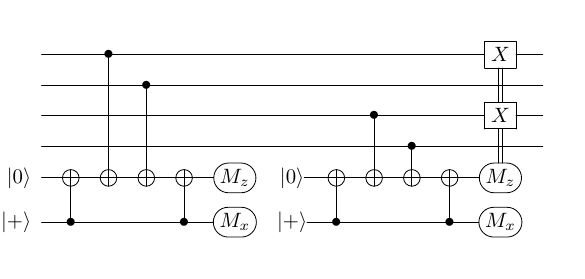}
    \caption{Fault-Tolerant $Z_1$ measurement in the \squarecode\ code with conditional $X$ corrections. No error is detected if both $X$ measurements return 0 and both $Z$ measurements return the same result. For the teleportation scheme the conditional $X$corrections are also applied to the qubits in the \cubecode\ code.}
    \label{fig:ft_Z}
\end{figure}

The first redundancy involves measuring multiple equivalent versions of the relevant logical operator to account for errors in the gates, preparations, and measurements, and then comparing the outcomes and decoding them.  
For these distance-two codes, it suffices to measure the logical operator just twice and accepting only if they both have the value of $+1$: first by non-destructively measuring logical $X$ or $Z$, and then by non-destructively measuring $SX$ or $SZ$ for a stabilizer element times that logical operator.  
Figure~\ref{fig:ft_X} depicts an example of this for the \cubecode\ code for measuring the logical $X_1$ then $S_5X$ operators as described in table~\ref{tab:codes}.  Figure~\ref{fig:ft_Z} depicts an example of this for the \squarecode\ code for measuring the logical $S_2Z$ then $Z_2$ operators as described in table~\ref{tab:codes}.  Geometrically, these measurements correspond to measuring operators supported by opposing faces or edges of the corresponding cube or square.

The second redundancy involves introducing an extra ``flag'' physical qubit that checks that the ancilla qubit has not suffered an error, as it is not protected by any code.  
In the original flag-fault-tolerant circuits for syndrome extraction, the coupling gates between the ancilla and flag qubits occurs after some of the gates coupling the ancilla and code block physical qubits.  
This is because if a flip of the ancilla qubit propagates to a stabilizer generator, it should not be flagged as a fault.
However, in the case of measuring a logical Pauli operator, it \emph{would} be a fault to propagate a Pauli error on the ancilla to a logical error on the code block.  
For this reason, the $\CNOT$ coupling gates between the flag and ancilla qubits straddle the ancilla-coupled non-destructive Pauli measurements.  
Examples of this are depicted in Figures~\ref{fig:ft_X} and \ref{fig:ft_Z}.

\section{Catalyzing the \texorpdfstring{$\{\textrm{Clifford}, T\}$}{\{Clifford, T\}} gate set from the \texorpdfstring{$\{CCZ, H, M_Z, \ket{0}, \ket{H}  \}$}{\{CCZ, H, MZ, |0>, |H>\}} gate set}
\label{sec:catalysis}

While it is impossible to exactly synthesize the $\{\textrm{Clifford}, T\}$ gate set from the $\{CCZ, H\}$ gate set, if we are allowed a single copy of a magic state, we can catalyze the $T$ gate using the $\{CCZ, H\}$ gate set.
The magic state we will consider is the $+1$ eigenstate of the Hadamard gate $\ket{H}$.\footnote{Other magic states could also be used to catalyze the $T$ gate.}

\begin{align}
  |H\> &:= \cos \frac{\pi}{8}|0\> + \sin \frac{\pi}{8}|1\>.
\end{align}

\begin{figure}[!ht]
    \centering
    \includegraphics{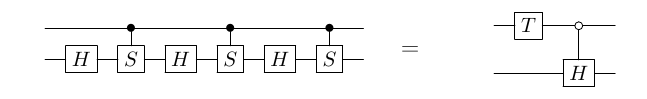}
    \caption{Controlling only $S$ gates in the sequence $SHSHSH$ performs a $|0\>$-controlled $H$ gate followed by a $T$ gate on the control qubit.}
    \label{fig:T_controlsequence}
\end{figure}

The gate sequence $SHSHSH$ is equivalent to the identity up to a global phase of $e^{i\pi/4}$.  
By applying this sequence controlled on the state of another qubit, this phase is turned into a relative phase between the $\ket{0}$ and $\ket{1}$ states of the control qubit. 
Recalling that the $T$ gate is $\Lambda(e^{i\pi/4})$ from Equation~(\ref{eq:Lambda-ZST}), this is just another representation of the $T$ gate.
If we can control this gate sequence by adding controls to just the $S$ gates, when the control qubit is in the state $|1\>$, the full sequence will be performed.
However, when the control qubit is in the state $|0\>$, only two Hadamard gates will cancel, leaving one net $H$ gate applied to the target qubit.  
This yields the circuit depicted in Figure~\ref{fig:T_controlsequence}. 

While we have no way of uncomputing this $|0\>$-controlled Hadamard gate, the action from this gate can be avoided if the state of the target qubit is the $+1$ eigenstate of the $H$ gate (\viz, the $\ket{H}$ state).

\begin{figure}[!ht]
    \centering
    \includegraphics{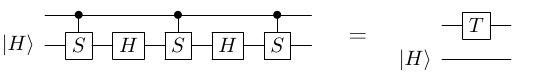}
    \caption{Controlling the sequence $SHSHSH$ with the
    target state set to the $\ket{H}$ state performs the $T$ gate catalytically.}
    \label{fig:T_noCH}
\end{figure}

Expanding into the phase-reference encoding, the catalytic circuit for exact synthesis of the $T$ gate using the $\{CCZ, H, |H\>\}$ gate set is the following:

\begin{figure}[!ht]
    \centering
    \includegraphics{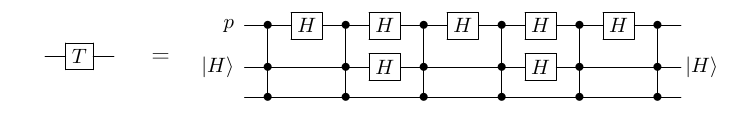}
    \caption{Exact synthesis of the $T$ gate using the $\{\CCZ, H, |H\>\}$ gate set catalytically in the phase-reference encoding.}
    \label{fig:T_CCZH}
\end{figure}

If we expand the gate set to include $\Lambda(Z)$, $\Lambda(X)$, $Z$, and $X$, the number of gates required for transpiling the $T$ gate can be reduced by only controlling the middle $S$ in the sequence $SHSHS$. 
Removing two controls will result in a $\ket{0}$-controlled $Z$ gate being applied, however, this gate is easily uncomputed.

\begin{figure}[!ht]
    \centering
    \includegraphics{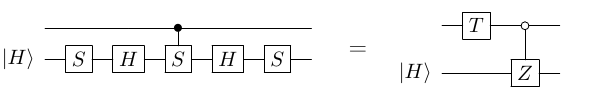}
    \caption{Enacting a $T$ gate by controlling only the central $S$ in the
    sequence $SHSHS$ also performs a low-controlled $Z$.}
    \label{fig:TCZ}
\end{figure}

\begin{figure}[!ht]
    \centering
    \includegraphics{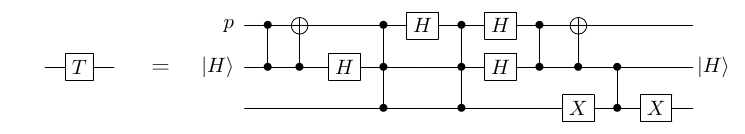}
    \caption{Decomposition of the $T$ gate into expanded gate set using $\ket{H}$
    catalytically. The phase-reference qubit is labeled $p$.}
    \label{fig:Tdecomp}
\end{figure}

The $S$ gates and single $\Lambda(S)$ can now be transpiled to give the decomposition of the $T$ gate in the expanded gate set. 
The $\ket{0}$-controlled $Z$ can be uncomputed with a $\Lambda(Z)$ conjugated by $X$ gates.
The transpiled $T$ gate is shown in Figure~\ref{fig:Tdecomp}.

The effect of infidelity on the $\ket{H}$ state can seen through calculating the action of the $\Lambda(H)$ gate on a catalyst state $\ket{\phi_H}=\sqrt{\gamma}\ket{H}+\sqrt{1-\gamma}\ket{-H}$

\begin{equation}
    \begin{split}
        \Lambda(H)\ket{\psi}\ket{\phi_H} = \sqrt{\gamma}\ket{\psi}\ket{H} + \sqrt{1-\gamma}(Z\ket{\psi})\ket{-H}
    \end{split}    
\end{equation}

We see that with probability $\gamma$ the phase kickback of the $\Lambda(H)$ is avoided and with probability $\left(1-\gamma\right)$ a logical $Z$ error occurs. 
The system and the catalyst become entangled, resulting in the same probability of success/error for all subsequent transpiled $T$ gates. 
In other words, with probability $\gamma$, all transpiled $T$ gates are performed correctly, and with probability $\left(1-\gamma\right)$, all transpiled $T$ gates suffer an error.
Thus, regardless of circuit depth, using a catalyst with infidelity $\left(1-\gamma\right)$ contributes a constant $\left(1-\gamma\right)$ to the overall error rate.

\section{Mirror circuits and confidence estimation}
\label{sec:C_est}

In this appendix we go into more detail on our benchmarking protocol proposed in Section~\ref{sec:performance}.

As stated in the main text, our confidence estimation protocol is an adaptation of reference~\cite{proctor2022establishing}, as such, much of this section is covered there.
We start by defining a mirror circuit $M$ as a product of logical layers $L_i$

\begin{equation}
    M = L_1^{-1}L_2^{-1}...L_{d/2}^{-1}L_{d/2}..L_2L_1
\end{equation}

We have the requirement that each layer consists only of encoded logical gates, namely, that each layer is contained within the normalizer of the QEC code stabilizer group. 

\begin{equation}
    L_i \in N(S)
\end{equation}

We model the noisy implementation of each layer as

\begin{equation}
    \phi(L_i) = \mathcal{E}(L_i)\mathcal{U}(L_i)
\end{equation}

Where $\mathcal{E}(L_i)$ and $\mathcal{U}(L_i)$ are superoperators and represent the noise associated with $L_i$ and the ideal unitary implementation of the layer respectively.
The noisy implementation of the circuit $M$ is then

\begin{equation}
    \phi(M) = \mathcal{E}(L_1^{-1})\mathcal{U}(L_1^{-1})...\mathcal{E}(L_{d/2}^{-1})\mathcal{U}(L_{d/2}^{-1})\mathcal{E}(L_{d/2})\mathcal{U}(L_{d/2})...\mathcal{E}(L_{1})\mathcal{U}(L_{1})
\end{equation}

We can rewrite $\phi(M)$ by commuting all the error terms to the right side

\begin{equation}
    \phi(M) = \mathcal{U}(L_1^{-1})...\mathcal{U}(L_{d/2}^{-1})\mathcal{U}(L_{d/2})...\mathcal{U}(L_{1})\mathcal{E'}(L_1^{-1})...\mathcal{E'}(L_{d/2}^{-1})\mathcal{E'}(L_{d/2})...\mathcal{E'}(L_{1})
\end{equation}

Where each $\mathcal{E'}(L_i)$ is the evolution of the error in layer $L_i$ through the rest of the circuit.

\begin{equation}
    \mathcal{E'}(L_i) = \mathcal{U}^{\dag}(L_i)...\mathcal{U}^{\dag}(L_1)\mathcal{E'}(L_i)\mathcal{U}^{\dag}(L_1)...\mathcal{U}^{\dag}(L_i)
\end{equation}

By construction, $\mathcal{U}(L_1^{-1})...\mathcal{U}(L_{d/2}^{-1})\mathcal{U}(L_{d/2})...\mathcal{U}(L_{1}) = \mathbb{1}$, so the noisy implementation of $M$ can be written purely as an error channel.

\begin{equation}
    \phi(M) = \mathcal{E'}(L_1^{-1})...\mathcal{E'}(L_{d/2}^{-1})\mathcal{E'}(L_{d/2})...\mathcal{E'}(L_{1}) = \mathcal{E}_{eff}
\end{equation}

Now we are utilizing error detection and post-selection which modifies the error channel as

\begin{equation}
    \phi_{ps}(M) = \Tilde{\Pi}_{ps}\mathcal{E}_{eff} = \Tilde{\mathcal{E}}_{ps}
\end{equation}

Where we have seen from the previous section that if the error detection is perfect $\Tilde{\mathcal{E}}_{ps}$ will be a stochastic logical Pauli channel.
Outside of that ideal limit $\Tilde{\mathcal{E}}_{ps}$ looks like the sum of the stochastic logical Pauli channel and a stochastic channel with Pauli's outside of the logical subspace.
Regardless of $\Tilde{\mathcal{E}}_{ps}$, however, a single logical qubit Clifford twirl will transform the marginal logical error probabilities into single logical qubit depolarizing channels~\cite{Gambetta_2012}.
This can be seen by looking at the marginal error probability of a single qubit

\begin{equation}
    \mathcal{E}_{marg} = p_0\mathbb{1} + p_x\mathcal{X} + p_y\mathcal{Y} + p_z\mathcal{Z}
\end{equation}

By definition, Clifford gates transform the Pauli operators into Pauli operators, therefore when we twirl $\mathcal{E}_{marg}$ by each Pauli superoperator gets transformed into uniform combination of all three operators and the likelihood of $\mathcal{X}$, $\mathcal{Y}$, and $\mathcal{Z}$ become equal regardless of $p_x$, $p_y$, and $p_z$.

\begin{equation}
    \frac{1}{|C_1|}\sum_{L_c \in C_1} L_c^{\dag} \mathcal{E}_{marg} L_c = p_0\mathbb{1} + \frac{p_x+p_y+p_z}{3}\left( \mathcal{X} + \mathcal{Y} + \mathcal{Z} \right)
    \label{eq:twirl}
\end{equation}

We see performing this twirl indeed transforms $\mathcal{E}_{marg}$ into a single-qubit logical depolarizing channel.
The twirl is performed in practice by applying a different random single qubit Clifford layer $L_c$ at the beginning and it's inverse $L_c^{\dag}$ at the end of the circuit and averaging over the outcomes.
In our architecture, directly preparing $\ket{+}$ and $\ket{-}$ states is easier than performing Hadamard gates, we therefore absorb the Clifford layer into the state preparation and measurement.

If there is an error on a single logical qubit that there is a $2/3$ probability that the bit is flipped when the qubit is measured.
This can be seen because Equation~(\ref{eq:twirl}) implies $\mathcal{X}$, $\mathcal{Y}$, and $\mathcal{Z}$ are equally likely but only $\mathcal{X}$ and $\mathcal{Y}$ cause the bit to flip in the Z-basis.
Equivalently, we when absorb the Clifford layer into the state preparation and measurement it randomizes the basis in which the qubit will be measured and $\mathcal{X}$, $\mathcal{Y}$, and $\mathcal{Z}$ will each cause a bit flip in $2$ of the $3$ bases.
More generally, on $n$ qubits a weight $W$ error will cause $k$ bitflips with probability $A_{kW}$ where by simple combinatorics  

\begin{equation}
     A_{kW} = \binom{W}{k}\frac{2^k}{3^W}
\end{equation}

We can then relate the probability that a weight $W$ error ($p_W$) occurs to the probability ($h_k$) that the measured output of circuit $M$ has a Hamming distance $k$ away from its ideal output

\begin{equation}
    h_k = A_{kW}p_W
    \label{eq:hk}
\end{equation}

The confidence for circuit $M$ ($C(M)$) is defined as the probability that there are no errors in a run after the run has been accepted by the post-selection.
As the post-selection has been included in the error channel $\mathcal{E}_{ps}$, we see $C(M)$ is simply the probability that $\mathcal{E}_{ps}$ causes a weight $0$ error (\emph{e.g.,} no error).
Therefore $C(M) = p_0$.
We obtain an estimate of $C(M)$ by inverting Equation~(\ref{eq:hk}) and summing over the $h_k$'s

\begin{equation}
    C(M) = \sum_k^n \left( -\frac{1}{2} \right)^{k}h_k
\end{equation}

This gives an estimate of the confidence for circuit $M$.
If we are interested in the average confidence for circuits of width $w$ and depth $d$, we can construct a set of circuits $\{M_i(w,d) \}$, estimate the confidence of each circuit in the set, then return the average measured confidence,

\begin{equation}
    \overline{C}(w,d) = \frac{1}{|\{M_i\}|}\sum_i C(M_i).
\end{equation}

\end{document}